\documentclass[11pt,a4paper]{article} 
\usepackage{bm}
\usepackage{latexsym}
\usepackage{amssymb}
\usepackage{amsmath}
\usepackage{multicol}
\usepackage{amsthm}
\usepackage[breaklinks]{hyperref} 
\usepackage{verbatim}
\usepackage{moreverb}
\usepackage{graphicx}
\usepackage{color}

\newtheorem{proposition}{Proposition}
\newtheorem{definition}{Definition}

\newtheorem{example}{Example}

\newcommand{\set}[2]{\{ \, #1 \mid #2 \, \}}

\renewcommand{\epsilon}{\varepsilon}
\renewcommand{\And}{\mathop{\&}}

\definecolor{aocolour}{rgb}{0.7,0.8,1}

\definecolor{grgreen}{cmyk}{1,0,1,0.6}
\definecolor{grred}{cmyk}{0,1,1,0.4}
\definecolor{grlmaroon}{cmyk}{0,0.6,0.6,0.6}
\definecolor{grnavy}{cmyk}{1,1,0,0.12}
\definecolor{grdarkcyan}{cmyk}{1,0,0,0.7}
\newcommand{\grt}[1]{\texttt{\textcolor{grred}{#1}}}

\newcommand{\gra}[1]{\emph{\textcolor{grnavy}{#1}}}
\newcommand{\grspace}{{\Large\textvisiblespace}}

\newcommand{\grsExpr}{E}
\newcommand{\grsExprCall}{E^{\mathrm{call}}}

\newcommand{\grsListOfExpr}{Z_{\mathrm{expr}}}
\newcommand{\grsListOfExprOnePlus}{Z_{\mathrm{expr}}^{1+}}
\newcommand{\grsListOfDistinctIds}{Z_{\mathrm{distinct\text{-}id}}}
\newcommand{\grsListOfDistinctIdsTwoPlus}{Z_{\mathrm{distinct\text{-}id}}^{2+}}

\newcommand{\grsStatement}{S}
\newcommand{\grsStatements}{S^*}
\newcommand{\grsStatementVar}{S^{\mathrm{var}}}
\newcommand{\grsStatementReturn}{S^{\mathrm{return}}}
\newcommand{\grsStatementRet}{S_r}
\newcommand{\grsStatementCompoundRet}{S^{\mathrm{compound}}_r}
\newcommand{\grsStatementNotIfThen}{S^{\mathrm{not\text{-}if\text{-}then}}}
\newcommand{\grsStatementNotVar}{S^{\mathrm{not\text{-}var}}}

\newcommand{\grsFunction}{F}
\newcommand{\grsFunctionMain}{F_{\mathrm{main}}}
\newcommand{\grsFunctionNotMain}{F_{\lnot \mathrm{main}}}
\newcommand{\grsFunctionHeader}{F_{\mathrm{header}}}
\newcommand{\grsFunctions}{F^*}
\newcommand{\grsFunctionsWithMain}{F^{*m*}}

\newcommand{\grsProgram}{\mathrm{Program}}

\newcommand{\anyletter}{\textup{anyletter}}
\newcommand{\anydigit}{\textup{anydigit}}
\newcommand{\anychar}{\textup{anychar}}
\newcommand{\anyletterdigit}{\textup{anyletterdigit}}
\newcommand{\anypunctuator}{\textup{anypunctuator}}
\newcommand{\anypunctuatorexceptrightpar}{\textup{anypunctuatorexceptrightpar}}
\newcommand{\anycharexceptbracespace}{\textup{anycharexceptbracesandspace}}
\newcommand{\anyletterdigits}{\textup{anyletterdigits}}
\newcommand{\anystring}{\textup{anystring}}
\newcommand{\anystringwithoutbraces}{\textup{anystringwithoutbraces}}
\newcommand{\anystringwithoutbracesandsemicolons}{\textup{anystringwithoutbracesandsemicolons}}
\newcommand{\safeendingstring}{\textup{safeendingstring}}
\newcommand{\returnstatementfix}{\textup{returnstatementfix}}
\newcommand{\Keyword}{\textup{Keyword}}

\newcommand{\newt}[1]{\text{\framebox{\grt{#1}}}}
\newcommand{\tComma}{\newt{,}}
\newcommand{\tSemicolon}{\newt{;}}
\newcommand{\tPlus}{\newt{$+$}}
\newcommand{\tMinus}{\newt{$-$}}
\newcommand{\tStar}{\newt{$*$}}
\newcommand{\tSlash}{\newt{$/$}}
\newcommand{\tMod}{\newt{mod}}
\newcommand{\tAnd}{\newt{$\land$}}
\newcommand{\tOr}{\newt{$\lor$}}
\newcommand{\tLessThan}{\newt{$<$}}
\newcommand{\tGreaterThan}{\newt{$>$}}
\newcommand{\tLessEqual}{\newt{$\leqslant$}}
\newcommand{\tGreaterEqual}{\newt{$\geqslant$}}
\newcommand{\tEqual}{\newt{$=$}}
\newcommand{\tNotEqual}{\newt{$\neq$}}
\newcommand{\tNot}{\newt{$\lnot$}}
\newcommand{\tLeftPar}{\newt{(}}
\newcommand{\tRightPar}{\newt{)}}
\newcommand{\tLeftBrace}{\newt{\{}}
\newcommand{\tRightBrace}{\newt{\}}}
\newcommand{\tAssign}{\newt{$:=$}}
\newcommand{\tIf}{\newt{if}}
\newcommand{\tElse}{\newt{else}}
\newcommand{\tWhile}{\newt{while}}
\newcommand{\tVar}{\newt{var}}
\newcommand{\tReturn}{\newt{return}}
\newcommand{\WS}{\textsc{ws}}

\newcommand{\tId}{\newt{\textsc{id}}}
\newcommand{\tIdAux}{\textup{Id}_1}
\newcommand{\tIdNotMain}{\newt{\textsc{id}$_{\lnot main}$}}
\newcommand{\tNum}{\newt{\textsc{num}}}
\newcommand{\tNumAux}{\textup{Num}_1}

\newcommand{\grC}{\text{\textcolor{grdarkcyan}{$C$}}}
\newcommand{\grCneg}{\text{\textcolor{grdarkcyan}{$\widetilde{C}$}}}
\newcommand{\Cmid}{C_{mid}}
\newcommand{\Clen}{C_{len}}
\newcommand{\Clenlt}{\widetilde{C}_{len<}}
\newcommand{\Clengt}{\widetilde{C}_{len>}}
\newcommand{\Citerate}{C_{iterate}}
\newcommand{\Citerateneg}{\widetilde{C}_{iterate}}
\newcommand{\Cc}[1]{C_{\texttt{#1}}}

\begin{document}

\sloppy

\title{Describing the syntax of programming languages
	using conjunctive and Boolean grammars\thanks{%
	This work was supported by the Russian Science Foundation, project 18-11-00100.
	}}
\author{Alexander Okhotin\thanks{%
	Department of Mathematics and Computer Science,
	St.~Petersburg State University, 7/9 Universitetskaya nab., Saint Petersburg 199034, Russia,
	\texttt{alexander.okhotin@spbu.ru}.}}

\maketitle

\begin{abstract}
A classical result by Floyd 
	(\href{http://dx.doi.org/10.1145/368834.368898}
	{``On the non-existence of a phrase structure grammar for ALGOL 60''},
	1962)
states that the complete syntax
of any sensible programming language
cannot be described
by the ordinary kind of formal grammars (Chomsky's ``context-free'').
This paper uses grammars extended with conjunction and negation operators,
known as conjunctive grammars and Boolean grammars,
to describe the set of well-formed programs
in a simple typeless procedural programming language.
A complete Boolean grammar, which defines such concepts
as declaration of variables and functions before their use,
is constructed and explained.
Using the Generalized LR parsing algorithm for Boolean grammars,
a program can then be parsed in time $O(n^4)$ in its length,
while another known algorithm allows subcubic-time parsing.
Next, it is shown how to transform this grammar
to an unambiguous conjunctive grammar,
with square-time parsing.
This becomes apparently the first specification
of the syntax of a programming language
entirely by a computationally feasible formal grammar.
\end{abstract}

\tableofcontents

\section{Introduction}

Formal grammars emerged
in the early days of computer science,
when the development of the first programming languages
required mathematical specification of syntax.
The first and the most basic model,
independently introduced by Chomsky~\cite{Chomsky1959}
(as ``phrase structure grammars'', later ``context-free grammars''),
and in the Algol 60 report~\cite{Algol60RevisedReport}
(as ``metalinguistic formulae'', later ``Backus--Naur form''),
is universally recognized as the standard model of syntax,
the ordinary kind of formal grammars.

Already in the Algol 60 report,
a grammar was used to describe
the complete lexical composition of the language
and the basic elements of its syntax.
More complicated syntactical requirements,
such as the rules requiring declaration of variables before use,
were described in plain words
together with the semantics of the language.
The hope of inventing a more sophisticated grammar
that would describe the entire syntax of Algol 60
was buried by Floyd~\cite{Floyd},
who proved that no (ordinary) grammar 
could ensure declaration before use.
Floyd considered strings of the following form,
here translated from Algol 60 to C.
\begin{equation*}
	\texttt{main() \{ int $\underbrace{\texttt{x} \ldots \texttt{x}}_{i \geqslant 1}$;
		$\underbrace{\texttt{x} \ldots \texttt{x}}_{j \geqslant 1}$ =
		$\underbrace{\texttt{x} \ldots \texttt{x}}_{k \geqslant 1}$; \}}
\end{equation*}
Such a string is a valid program
	if and only if
$i=j=k$,
for otherwise the assignment statement would refer to an undeclared variable.
This very simple fragment of the programming language
is an instance of an abstract formal language
$L_1=\set{a^n b^n c^n}{n \geqslant 0}$,
and since it is known that no
ordinary formal grammar (that is, Chomsky's ``context-free grammar'')
could be constructed for the latter language,
one can infer that
no such grammar can describe the whole programming language.

Many other simple syntactic conditions
common to all programming languages
are also beyond the scope of ordinary formal grammars.
For instance, strings of the following form
are valid programs
	if and only if
$i=k$ and $j=\ell$,
because otherwise the number of arguments in one of the calls
would not match that in the function prototype.
\begin{equation*}
	\texttt{%
	void f($\underbrace{\texttt{int, $\ldots$, int}}_{i \geqslant 1}$);
	void g($\underbrace{\texttt{int, $\ldots$, int}}_{j \geqslant 1}$);
	main()
	\{
		f($\underbrace{\texttt{0, $\ldots$, 0}}_{k \geqslant 1}$);
		g($\underbrace{\texttt{0, $\ldots$, 0}}_{\ell \geqslant 1}$);
	\}}
\end{equation*}
This construct is modelled by the formal language
$L_2=\set{a^m b^n a^m c^n}{m,n \geqslant 0}$,
which is another standard example of a language
not described by any (ordinary) grammar.
The next example corresponds to mathing a pair of identifiers
over a two-symbol alphabet;
these strings,
defined for any two identifiers $w, w' \in \{\texttt{a}, \texttt{b}\}^+$,
are valid programs 
	if and only if
$w=w'$.
\begin{equation*}
	\texttt{main() \{
		int $w$;
		$w'$ = 0;
		\}}
\end{equation*}
This is an instance of the formal language
$L_3=\set{wcw}{w \in \{a,b\}^*}$,
investigated, in particular,
by Soko{\l}owski~\cite{Sokolowski}
and by Wotschke~\cite{Wotschke}.
Not only it is not described by any (ordinary) grammar,
it is not representable as an intersection
of finitely many languages described by grammars~\cite{Wotschke}.

Floyd's~\cite{Floyd} result,
besides pointing out important limitations of grammars,
can also be regarded as a formulation of a problem:
that of defining an extended formal grammar model
powerful enough to describe those constructs of programming languages
that are beyond the power of ordinary grammars.
Viewed in this way, the result of Floyd
has prompted formal language theorists
to search for such a model.
Many early attempts ended up with seemingly promising new grammar families,
which, at a closer examination,
turned out to be too powerful
to the point of being useless:
this can be said of Chomsky's~\cite{Chomsky1959} ``context-sensitive grammars''
and of van Wijngaarden's~\cite{vanWijngaarden} ``two-level grammars'',
which can simulate Turing machines.
Accordingly, these models define \emph{computations},
rather than any kind of syntactic structures.
Even at their best, any grammars defined in these models
are actually \emph{parsers} that are executed as programs,
rather than parser-independent specifications of the syntax;
as such, they are beyond any formal analysis.

Besides these unsuccessful attempts,
a few generalized grammar models
capable of giving meaningful descriptions of syntax
were discovered.
The first such models were
Aho's \emph{indexed grammars}~\cite{AhoIndexed},
Fischer's \emph{macro grammars}~\cite{Fischer_macro}
and \emph{tree-adjoining grammars} by Joshi et al.~\cite{JoshiLevyTakahashi}.
Later, the ideas behind these models
led to the more practical
\emph{multi-component grammars}~\cite{SekiMatsumuraFujiiKasami,VijayShankerWeirJoshi},
which became a standard model in computational linguistics
and receive continued attention.

However, even though these models are powerful enough
to define the above three abstract languages $L_1$, $L_2$ and $L_3$,
the mere existence of grammars for these languages
does not at all imply
that a grammar for any programming language can be constructed.
On the contrary, a simple extension of Floyd's example
shows that, again, no multi-component grammar
can describe the set of well-formed programs.
Consider the following strings,
in which the same variable
is referenced an unbounded number of times.
\begin{equation*}
	\texttt{main() \{ int $\underbrace{\texttt{x} \ldots \texttt{x}}_{i \geqslant 1}$;
		$\underbrace{\underbrace{\texttt{x} \ldots \texttt{x}}_{j_1 \geqslant 1} = \texttt{0;}
		\;\ldots\;
		\underbrace{\texttt{x} \ldots \texttt{x}}_{j_k \geqslant 1} = \texttt{0;}}_{k \geqslant 0}$
	\}}
\end{equation*}
Such a string is a well-formed program
	if and only if
all numbers $j_1, \ldots, j_k$
are equal to $i$.
This is an instance of an abstract language
$L_4=\set{(a^n b)^k}{n \geqslant 0, \: k \geqslant 1}$,
which is known to have no multi-component grammar,
because its commutative image is not a semilinear set~\cite[Sect.~4.2]{VijayShankerWeirJoshi}.

It is still possible to construct an indexed grammar for $L_4$.
However, it is not difficult to present
yet another simple case of programming language syntax
that is beyond their expressive power.
This time, consider strings of the following form.
\begin{equation*}
	\texttt{%
	int f($\underbrace{\texttt{int, $\ldots$, int}}_{i \geqslant 1}$);
	main()
	\{
		f($\underbrace{\texttt{f(}\underbrace{\texttt{0,} \ldots \texttt{,0}}_{j_1 \geqslant 1}\texttt{),}
		\;\ldots\;
		\texttt{, f(}\underbrace{\texttt{0,} \ldots \texttt{,0}}_{j_k \geqslant 1}
		\texttt{)}}_{k \geqslant 1}$);
	\}}
\end{equation*}
Such a string is a well-formed program
	if and only if
$i=k=j_1=\ldots=j_k$.
This is an instance of an abstract language
$L_5=\set{(a^n b)^n}{n \geqslant 1}$,
which cannot be described by an indexed grammar~\cite{Gilman}.

Even though abstract languages, such as these,
have occasionally been brought forward
to claim practical relevance of new grammar models,
they are not representative
of the syntactic constructs in programming languages.
In order to show that some kind of formal grammars
are powerful enough to define those syntactic constructs,
the only convincing demonstration
would be a complete grammar
for some programming language.
This paper provides such a demonstration
for the families of \emph{conjunctive grammars}
and \emph{Boolean grammars},
constructing a complete grammar
for the set of well-formed programs
in a simple model procedural language featuring a single data type,
a standard set of flow control statements and
nested compound statements with rules for variable scope.

Conjunctive grammars~\cite{Conjunctive,BooleanSurvey}
extend Chomsky's ``context-free'' grammars
by allowing a conjunction of any syntactic conditions
to be expressed in any rule.
Consider that a rule $A \to BC$ in an ordinary grammar
states that if a string $w$ is representable as $BC$---%
that is, as $w=uv$,
where $u$ has the property $B$ and $v$ has the property $C$---%
then $w$ has the property $A$.
In a conjunctive grammar, one can define a rule of the form
$A \to BC \And DE$,
which asserts that every string $w$
representable \emph{both} as $BC$ (with $w=uv$)
\emph{and at the same time} as $DE$ (with $w=xy$)
therefore has the property $A$.
The more general family of Boolean grammars~\cite{BooleanGrammars,BooleanSurvey}
further allows negation:
a rule $A \to BC \And \lnot DE$
states that if a string
is representable as $BC$ (with $w=uv$),
\emph{but is not representable} as $DE$,
then it has the property $A$.
In this way, the properties of a string
are defined independently of the context in which it occurs,
in the same way as in Chomsky's ``context-free'' grammars.
These models differ only is the set of allowed Boolean operations,
and for that reason,
the familiar kind of grammars featuring disjunction only
shall be called \emph{ordinary grammars} throughout this paper.

Even though conjunctive grammars have a Chomsky-like definition by term rewriting,
whereas Boolean grammars are defined by language equations
generalizing those by Ginsburg and Rice~\cite{GinsburgRice},
their true meaning lies in logic.
The understanding of ordinary grammars
in terms of logical inference
can be found, for instance,
in Kowalski's~\cite[Ch.~3]{Kowalski} textbook.
In one of the first papers
exploring more powerful logics inspired by grammars,
Schuler~\cite{Schuler_programming}
argues that the set of well-formed programs is Algol 60
can be described in his formalism~\cite{Schuler_inductive}.
For the modern logical understanding of grammars,
the reader is referred to an important paper by Rounds~\cite{Rounds},
who explained different kinds of formal grammars
as fragments of the FO(LFP) logic~\cite{Immerman_polynomial,Vardi_polynomial}.
Conjunctive grammars are another such fragment.

Conjunctive and Boolean grammars
are important for two reasons.
First, they enrich standard inductive definitions of syntax
with important logical operations,
which extend the expressive power of such definitions
in a practically useful way;
this shall be further supported in the present paper.
At the same time,
these grammars have generally the same parsing algorithms
as ordinary grammars~\cite{AizikowitzKaminski_LR0,BooleanGrammars,BooleanLR,BooleanLL,BooleanUnambiguous},
and share the same subcubic upper bound
on the time complexity of parsing~\cite{BooleanMatrix},
which makes them suitable for implementation.
Based on these properties,
Stevenson and Cordy~\cite{StevensonCordy}
recently applied Boolean grammars
to \emph{agile parsing} in software engineering.

The most practical parsing algorithm for Boolean grammars
is a variant of the Generalized LR (GLR)~\cite{BooleanLR},
which runs in worst-case time $O(n^4)$
and operates very similarly
to the GLR for ordinary grammars~\cite{Tomita}.
Two implementations of Boolean GLR are known~\cite{Megacz,WhaleCalf}.
In the literature,
GLR parsers have sometimes been applied
to analyzing programming languages symbol by symbol,
without an intermediate layer of lexical analysis~%
\cite{VandenbrandScheerderVinjuVisser,EconomopoulosKlintVinju,KatsDejongeNilssonnymanVisser}:
this is known as \emph{scannerless parsing}~\cite{SalomonCormack}.
The Boolean grammar for a programming language
constructed in this paper
follows the same principle,
and a Boolean GLR parser for the new grammar
is not much different from the GLR operating on an ordinary grammar.

The theoretical work on conjunctive and Boolean grammars
is reviewed in a recent survey paper~\cite{BooleanSurvey}.
This paper includes all the necessary definitions,
given in Section~\ref{section_conjunctive_boolean},
and illustrates the use of conjunction and negation
on two examples presented in Section~\ref{section_examples}.

The model programming language is defined
in Section~\ref{section_the_grammar},
and every point of the definition
is immediately expressed in the formalism of Boolean grammars.
The grammar is designed for scannerless parsing,
and for this reason alone,
it is bound to be somewhat involved,
with definitions of nested syntactic structure
occasionally interleaved with simulation of finite automata.
In the literature, this was metaphorically described
as skipping ``water'' in search for ``islands''~\cite{KatsDejongeNilssonnymanVisser}.
In a few places,
defining separation into tokens within the grammar
results in rather awkward rules.
This, however, is a general trait of scannerless parsing
and not a fault of Boolean grammars.

A certain improvement to this grammar
is presented in Section~\ref{section_unambconj},
where it is shown how to eliminate negation and ambiguity in it.
In other words, a Boolean grammar
is reformulated as an unambiguous conjunctive grammar~\cite{BooleanUnambiguous},
which is a conceptually easier model
with a better worst-case parsing complexity---%
namely, square time in the length of the input.
In the literature on scannerless parsing,
grammars are typically ambiguous,
with attached external disambiguation rules~%
\cite{BegelGraham,VandenbrandScheerderVinjuVisser,SalomonCormack}.
From this perspective, this section
demonstrates a new, entirely grammatical approach
to disambiguating scannerless parsers
for programming languages.

The paper is concluded
with two kinds of research directions,
suggested in Section~\ref{section_afterthoughts}.
First, what kind of parsers
could handle this or similar grammars
in less than square time?
Second, what kind of new grammar models
could describe the syntax of programming languages
more conveniently?

\paragraph{Note:}
An earlier form of the Boolean grammar in Section~\ref{section_the_grammar}
was presented at the AFL 2005 conference
held in Dobog\'ok\H{o}, Hungary,  
and published in a local proceedings volume~\cite{Prog_AFL}.
This paper supercedes that preliminary report.

\section{Conjunctive and Boolean grammars} \label{section_conjunctive_boolean}

\subsection{Conjunctive grammars}

In ordinary formal grammars,
rules specify how substrings are concatenated to each other,
and one can define \emph{disjunction} of syntactic conditions
by writing multiple rules for a nonterminal symbol.
In \emph{conjunctive grammars}, this logic is extended
to allow conjunction within the same kind of definitions.

\begin{definition}[\cite{Conjunctive,BooleanSurvey}]
A conjunctive grammar
is a quadruple $G=(\Sigma, N, R, S)$, in which:
\begin{itemize}
\item
	$\Sigma$ is the \emph{alphabet}
	of the language being defined;
\item
	$N$ is a finite set of \emph{nonterminal symbols},
	each representing a property of strings defined within the grammar;
\item
	$R$ is a finite set of \emph{rules}, each of the form
	\begin{equation}\label{conjunctive_rule} 
		A \to \alpha_1 \And \ldots \And \alpha_m,
	\end{equation}
	where $A \in N$, $m \geqslant 1$ and
	$\alpha_1, \ldots, \alpha_m \in (\Sigma \cup N)^*$;
\item
	$S \in N$ is a symbol
	representing the property of being
	a syntactically well-formed sentence of the language
	(``the initial symbol'').
\end{itemize}
\end{definition}

Each concatenation $\alpha_i$
in a rule (\ref{conjunctive_rule})
is called a \emph{conjunct}.
If a grammar has a unique conjunct in every rule ($m=1$),
it is an \emph{ordinary grammar} (Chomsky's ``context-free'').
If every conjunct contains at most one nonterminal symbol
($\alpha_1, \ldots, \alpha_m \in \Sigma^* N \Sigma^* \cup \Sigma^*$),
a grammar is called \emph{linear conjunctive}.
Multiple rules for the same nonterminal symbol
may be presented in the usual notation,
such as $A \to \alpha_1 \And \ldots \And \alpha_m \ | \ \beta_1 \And \ldots \And \beta_n$,
etc.
As in ordinary grammars,
the vertical line is essentially disjunction.

Each rule (\ref{conjunctive_rule}) means that
any string representable as each concatenation $\alpha_i$
therefore has the property $A$.
This understanding can be equivalently formalized
by term rewriting~\cite{Conjunctive}
and by language equations~\cite{ConjEquations}.
Consider the former definition, 
which extends Chomsky's definition of ordinary grammars
by string rewriting,
using terms instead of strings.

\begin{definition}[\cite{Conjunctive}]\label{conjunctive_rewriting_definition}
Let $G=(\Sigma, N, R, S)$ be a conjunctive grammar,
and consider terms over concatenation and conjunction,
with symbols from $\Sigma \cup N$ and the empty string $\epsilon$ as atomic terms.
The relation of one-step rewriting on such terms
($\Longrightarrow$)
is defined as follows.
\begin{itemize}
\item
	Using a rule $A \to \alpha_1 \And \ldots \And \alpha_m \in R$,
	with $A \in N$,
	any atomic subterm $A$ of any term
	may be rewritten by the term on the right-hand side of the rule,
	enclosed in brackets.
	\begin{equation*}
		\ldots A \ldots
			\Longrightarrow
		\ldots (\alpha_1 \And \ldots \And \alpha_m) \ldots
	\end{equation*}
\item	A conjunction of several identical strings 
	may be rewritten to one such string.
	\begin{align*}
		\ldots (w \And \ldots \And w) \ldots
			\Longrightarrow
		\ldots w \ldots
		&& (w \in \Sigma^*)
	\end{align*}
\end{itemize}
The language defined by a term $\varphi$
is the set of all strings over $\Sigma$
obtained from it in a finite number of rewriting steps.
\begin{equation*}
	L_G(\varphi) = \set{w}{w \in \Sigma^*, \; \varphi \Longrightarrow^* w}
\end{equation*}
The language described by the grammar
is the language defined by its initial symbol.
\begin{equation*}
	L(G) = L_G(S) = \set{w}{w \in \Sigma^*, \; S \Longrightarrow^* w}
\end{equation*}
\end{definition}

\begin{figure}[t]
	\centerline{\includegraphics[scale=0.8]{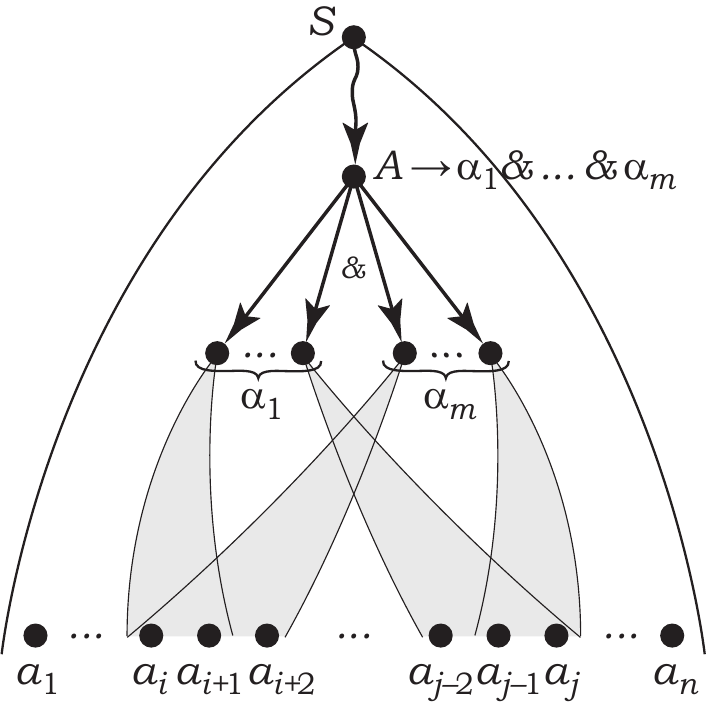}}
	\caption{Parse trees in conjunctive grammars:
		a subtree with root $A \to \alpha_1 \And \ldots \And \alpha_m$,
		representing $m$ parses of a substring $a_i \ldots a_j$.}
	\label{f:conjunctive_trees}
\end{figure}

An important property of conjunctive grammars
is that every string in $L(G)$
has a corresponding \emph{parse tree},
which is exactly a proof tree in this logic theory.
This is, strictly speaking, an acyclic graph
rather than a tree in a mathematical sense.
Its leaves (sinks) correspond to the symbols in $w$.
Every internal node in this tree
is labelled with some rule (\ref{conjunctive_rule}),
and has as many ordered children
as there are symbols in all conjuncts $\alpha_1$, \ldots, $\alpha_m$.
Subtrees corresponding to different conjuncts
in a rule define multiple interpretations of the same substring,
and accordingly lead to the same set of leaves,
as illustrated in Figure~\ref{f:conjunctive_trees}.
For succinctness, the label of an internal node
can be just a nonterminal symbol,
as long as the rule can be deduced
from the children's labels
(this is always the case in ordinary grammars,
but need not be true in conjunctive grammars).

\subsection{Boolean grammars}

The second family of grammars used in this paper
are \emph{Boolean grammars},
which further extend conjunctive grammars with a negation operator.
To be precise, negation can be put over any conjunct in any rule,
making it a \emph{negative conjunct}.

\begin{definition}[\cite{BooleanGrammars,BooleanSurvey}]
A Boolean grammar is a quadruple $G=(\Sigma, N, R, S)$,
where
\begin{itemize}
\item	$\Sigma$ is the alphabet;
\item	$N$ is the set of nonterminal symbols;
\item	$R$ is a finite set of rules of the form
	\begin{equation}\label{rule_in_Boolean_grammar} 
		A \to \alpha_1 \And \ldots \And \alpha_m \And \lnot\beta_1 \And \ldots \And \lnot\beta_n
	\end{equation}
	with $A \in N$, $m,n \geqslant 0$, $m+n \geqslant 1$ and
	$\alpha_i, \beta_j \in (\Sigma \cup N)^*$;
\item	$S \in N$ is the initial symbol.
\end{itemize}
\end{definition}

A conjunctive grammar is then a Boolean grammar,
in which every conjunct is positive.

A rule (\ref{rule_in_Boolean_grammar}) is meant to state that
every string
representable as each of $\alpha_1$, \ldots, $\alpha_m$,
but not representable as any of $\beta_1$, \ldots, $\beta_n$,
therefore has the property $A$.
This intuitive definition is formalized by using \emph{language equations},
that is, by representing a grammar
as a system of equations with formal languages as unknowns,
and using a solution of this system
as the language defined by the grammar.
The definition of Boolean grammars exists in two variants:
the simple one, given by the author~\cite{BooleanGrammars},
and the improved definition by Kountouriotis et al.~\cite{KountouriotisNomikosRondogiannis}
based on the well-founded semantics of negation in logic programming.
Even though the simple definition
handles some extreme cases of grammars improperly~\cite{KountouriotisNomikosRondogiannis},
it ultimately defines the same family of languages,
and is therefore sufficient in this paper.

For every integer $\ell \geqslant 0$,
let $\Sigma^{\leqslant \ell}$ denote the set of all strings over $\Sigma$
of length at most $\ell$.
With an alphabet $\Sigma$ fixed,
the \emph{complement} of a language $L \subseteq \Sigma^*$
is the language $\overline{L} = \set{w}{w \in \Sigma^*, \: w \notin L}$.

\begin{definition}[Okhotin~\cite{BooleanGrammars}]\label{s1s_definition}
Let $G=(\Sigma, N, R, S)$ be a Boolean grammar,
and consider the following system of equations,
in which every symbol $A \in N$ is an unknown language over $\Sigma$.
\begin{equation}\label{system_associated_to_Boolean_grammar}
	A = \bigcup_{\substack{A \to \alpha_1 \And \ldots \And \alpha_m \And \\ \And \lnot \beta_1 \And \ldots \And \lnot \beta_n \in R}}
		\bigg[
			\bigcap_{i=1}^m \alpha_i
			\;\cap\;
			\bigcap_{j=1}^n \overline{\beta_j}
		\bigg]
\end{equation}
Each symbol $B \in N$ used in the right-hand side of any equation
is a reference to a variable,
and each symbol $a \in \Sigma$ represents a constant language $\{a\}$.

Assume that, for every $\ell \geqslant 0$,
there exists a unique vector of languages
$(\ldots, L_A, \ldots)_{A \in N}$ with $L_A \subseteq \Sigma^{\leqslant \ell}$,
for which a substitution of $L_A$ for $A$, for all $A \in N$,
turns every equation (\ref{system_associated_to_Boolean_grammar})
into an equality modulo intersection with $\Sigma^{\leqslant \ell}$.
Then the system is said to have a \emph{strongly unique solution},
and, for every $A \in N$,
the language $L_G(A)$ is defined as $L_A$
from the unique solution of this system.
The language described by the grammar is $L(G)=L_G(S)$.
\end{definition}

If, for some $\ell \geqslant 0$,
the solution modulo $\Sigma^{\leqslant \ell}$ is not unique,
then the grammar is considered invalid.

Boolean grammars also define parse trees,
which, however, reflect only positive components of a parse~\cite{BooleanGrammars}.
An internal node labelled with a rule (\ref{rule_in_Boolean_grammar})
has children corresponding to the symbols
in the positive conjuncts $\alpha_1, \ldots, \alpha_m$,
which represent multiple parses of that substring,
exactly like in a conjunctive grammar.
Negative conjuncts have no representation in the tree.

\subsection{Ambiguity}

Informally, a grammar is unambiguous
if every string can be parsed in a unique way.
For ordinary grammars,
this is formalized by uniqueness of a parse tree.
For conjunctive grammars,
the same kind of definition
would no longer be useful,
because a grammar may define multiple parses
for some substrings,
only to eliminate those substrings later
using intersection:
in that case, the parse tree can still be unique,
but in terms of complexity of parsing,
such grammars are ambiguous.
For Boolean grammars,
a parse tree represents only partial information on the parse,
and a definition of ambiguity by parse tree uniqueness
becomes completely wrong.

These observations led to the following definition.

\begin{definition}[\cite{BooleanUnambiguous}] \label{boolean_unambiguity_definition}
A Boolean grammar $G=(\Sigma, N, R, S)$ is unambiguous, if
\begin{enumerate}\renewcommand{\theenumi}{\Roman{enumi}}
\item
	the choice of a rule for every single nonterminal $A$
	is unambiguous,
	in the sense that for every string $w$,
	there exists at most one rule
	\begin{equation*}
		A \to \alpha_1 \And \ldots \And \alpha_m \And \lnot\beta_1 \And \ldots \And \lnot\beta_n,
	\end{equation*}
	with
	$w \in L_G(\alpha_1) \cap \ldots \cap L_G(\alpha_m) \cap
		\overline{L_G(\beta_1)} \cap \ldots \cap \overline{L_G(\beta_n)}$
	(in other words, different rules generate disjoint languages),
	and
\item
	all concatenations are unambiguous,
	that is, for every conjunct $X_1 \ldots X_\ell$ or $\lnot X_1 \ldots X_\ell$
	that occurs in the grammar,
	and for every string $w$,
	there exists at most one partition
	$w=u_1 \ldots u_\ell$ with
	$u_i \in L_G(X_i)$ for all $i$.
\end{enumerate}
\end{definition}

The concatenation unambiguity requirement
applies to positive and negative conjuncts alike.
For a positive conjunct belonging to some rule,
this means that a string
that is potentially generated by this rule
must be uniquely split according to this conjunct.
For a negative conjunct $\lnot DE$,
this condition requests
that a partition of $w \in L_G(DE)$
into $L_G(D) \cdot L_G(E)$ is unique,
even though $w$ is \emph{not defined}
by any rule involving this conjunct.

\section{Language specification with conjunctive and Boolean grammars}\label{section_examples}

All five abstract languages mentioned in the introduction
are defined by conjunctive grammars:
grammars for the first three,
$L_1=\set{a^n b^n c^n}{n \geqslant 0}$,
$L_2=\set{a^m b^n a^m c^n}{m,n \geqslant 0}$
and
$L_3=\set{wcw}{w \in \{a,b\}^*}$,
appeared in the literature~\cite{Conjunctive,BooleanSurvey},
and grammars for the languages
$L_4=\set{(a^n b)^k}{n \geqslant 0, \: k \geqslant 1}$
and $L_5=\set{(a^n b)^n}{n \geqslant 1}$
are given in Examples~\ref{anb_k_example}--\ref{anb_n_example} below.
What is more important,
is that these syntactical elements of programming languages
can be described in such a way
that generalizes further to a grammar
for an entire programming language.

A conjunctive grammar for $L_4$ is given below.
In this language, all strings of the form $(a^n b)^k$,
with $n \geqslant 0$ and $k \geqslant 1$,
model $k-1$ references to the same declaration $a^n$.
In an ordinary grammar, one can define such strings only for $k=2$,
and a typical grammar will use a rule $C \to aCa$
to match the number of symbols $a$ in the first block (declaration)
to that in the second block (reference).
The following grammar arranges the same matching
to be done between the first block
and every subsequent block.

\begin{example}\label{anb_k_example}
The following conjunctive grammar describes the language
$L_4=\set{(a^n b)^k}{n \geqslant 0, \: k \geqslant 1}$.
\begin{equation*}\begin{array}{rcl}
	S &\to& SA \And Cb \ | \ A \\
	A &\to& aA \ | \ b \\
	C &\to& aCa \ | \ B \\
	B &\to& BA \ | \ b
\end{array}\end{equation*}
\end{example}

The rules for $A$ and for $B$
define regular languages $L(A)=a^*b$
and $L(B)=b (a^*b)^*$.
Then, $C$ defines the language
$L(C)=\set{a^n x a^n}{n \geqslant 0, \: x \in b (a^*b)^*}$
representing a single identifier check,
carried out in the standard way.
The rules for $S$ arrange for all references
to be checked by $C$.

\begin{figure}[t]
	\centerline{\includegraphics[scale=0.8]{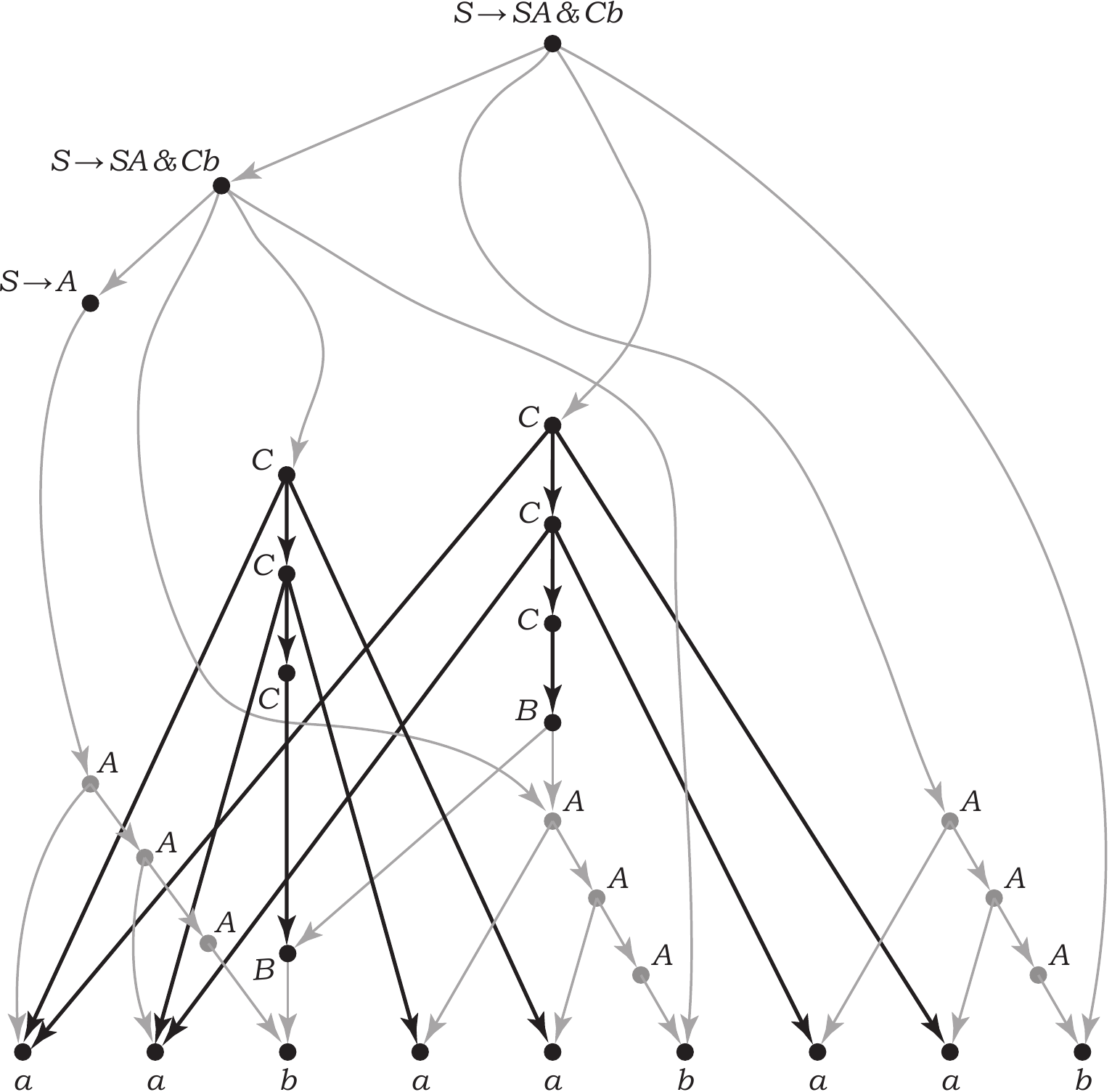}}
	\caption{Parse tree of the string $aabaabaab$
		for the grammar in Example~\ref{anb_k_example}.}
	\label{f:anb_k_parse_tree}
\end{figure}

All strings $(a^n b)^k$
are defined by $S$
inductively on $k$.
If $k=1$, then a string $a^n b$,
representing a lone declaration without references,
is given by a rule $S \to A$.
For $k \geqslant 2$,
the rule $S \to SA \And Cb$
imposes two conditions on a string.
First, the conjunct $SA$
declares the string to be a concatenation
of a string $(a^n b)^{k-1}$
with any string $a^m b \in L(A)$;
this verifies that all earlier references are correct.
The other conjunct $Cb$
compares the number of symbols $a$
in the first block $a^n b$ (the declaration)
to that in the last block $a^m b$ (the last reference).
This ensures that the string
is actually $(a^n b)^k$, as desired.

The parse tree of the string $aabaabaab$
is given in Figure~\ref{f:anb_k_parse_tree}.
The parts emphasized by thick black lines show $C$
comparing the number of symbols $a$
in the first block
to that in the second block
and in the third block.
The nodes in the upper part of the tree,
labelled with different rules for $S$,
arrange these comparisons to be made.

\begin{figure}[t]
	\centerline{\includegraphics[scale=0.8]{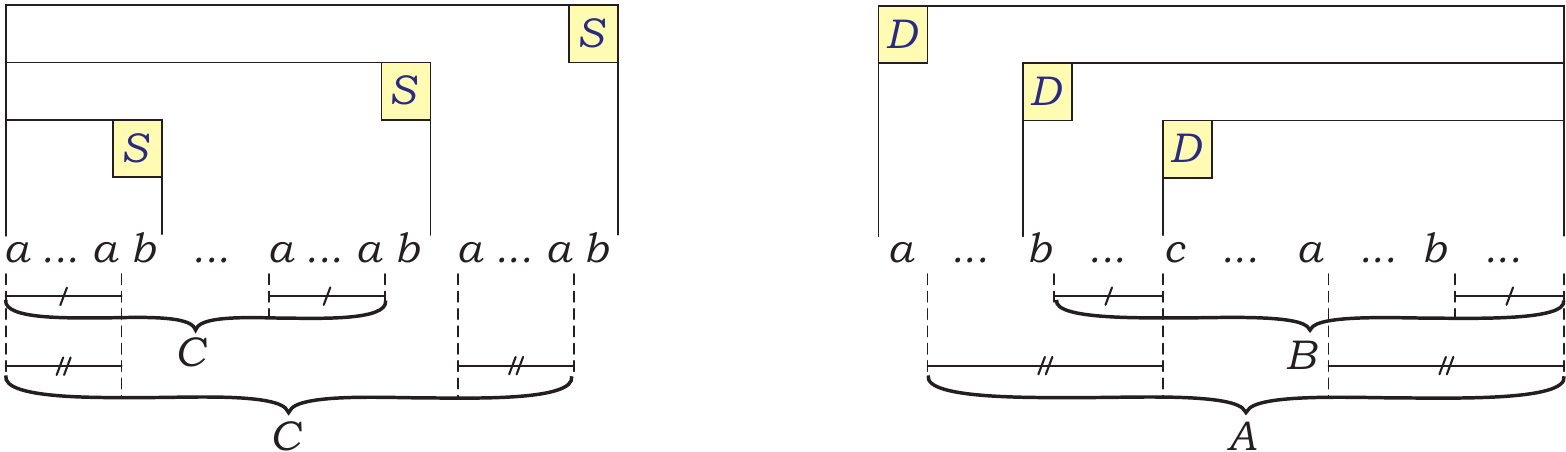}}
	\caption{(left) How the grammar in Example~\ref{anb_k_example}
		defines strings of the form $(a^n b)^k$;
		(right) How the grammar in Example~\ref{wcw_example}
		defines strings of the form $uczu$.}
	\label{f:anb_k_wcw_diagrams}
\end{figure}

Alternatively, the structure of comparisons
defined in this grammar
is illustrated in the informal diagram
in Figure~\ref{f:anb_k_wcw_diagrams}(left),
where the upper part shows how the rule $S \to SA \And Cb$
recursively refers to $S$ for shorter substrings.
The lower part of the diagram
illustrates the length equality defined by $C$.
All subsequent grammars in this paper
shall be illustrated by similar diagrams.

A grammar for $L_5$ can be obtained by reusing the grammar for $L_4$ in the following way.
\begin{example}\label{anb_n_example}
The language $L_5=\set{(a^n b)^n}{n \geqslant 1}$ is an intersection of $L_4$
with the language $L'=\set{a^n b (a^*b)^n}{n \geqslant 1}$.
A grammar for the latter language is not difficult to construct.
Then it remains to combine the grammars for $L_4$ and for $L'$
with a conjunction operator.
\end{example}

Even though the language in Example~\ref{anb_k_example}
is just one simple abstract language,
the grammar construction technique for conjunctive grammars
demonstrated in this example
is sufficient to arrange all identifier checks
in a simple programming language.
Another essential element
is the ability to compare identifiers
over a multiple-symbol alphabet,
which is modelled in the next example.

\begin{example}[Okhotin~\cite{Conjunctive}]\label{wcw_example}
The following conjunctive grammar
describes the language $\set{wcw}{w \in \{a, b\}^*}$.
\begin{equation*}\begin{array}{rcl}
	S &\to& C \And D \\
	C &\to& XCX \ | \ c \\
	D &\to& aA \And aD \ | \ bB \And bD \ | \ cE \\
	A &\to& XAX \ | \ cEa \\
	B &\to& XBX \ | \ cEb \\
	E &\to& XE \ | \ \epsilon \\
	X &\to& a \ | \ b
\end{array}\end{equation*}
\end{example}

First, $C$ defines the language of all strings $xcy$,
with $x, y \in \{a, b\}^*$ and $|x|=|y|$,
and thus the conjunction with $C$ in the rule for $S$ ensures
that the string consists of two parts of equal length
separated by a center marker.
The other conjunct $D$ checks
that the symbols in corresponding positions are the same.
The actual language defined by $D$
is $L(D)=\set{uczu}{u, z \in \{a, b\}^*}$,
and the rules for $D$ define these strings inductively as follows:
a string is in $L(D)$
	if and only if
\begin{itemize}
\item
	either it is in $c \{a,b\}^*$ (the base case: no symbols to compare),
\item
	or its first symbol is the same
	as the corresponding symbol on the other side,
	\emph{and} the string without its first symbol is in $L(D)$
	(that is, the rest of the symbols in the left part
	correctly correspond to the symbols in the right part).
\end{itemize}
The comparison of a single symbol to the corresponding symbol on the right
is done by the nonterminals $A$ and $B$,
which generate the languages $\set{xcvay}{x, v, y \in \{a, b\}^*, \: |x|=|y|}$
and $\set{xcvby}{x, v, y \in \{a, b\}^*, \: |x|=|y|}$, respectively,
and the above inductive definition
is directly expressed in the rules for $D$,
which recursively refer to $D$
in order to ensure that the same condition
holds for the rest of the string.

The grammar in Example~\ref{wcw_example}
essentially relies on having a center marker $c$
between the first and the second $w$.
The same language without a center marker,
$\set{ww}{w \in \{a, b\}^*}$,
cannot be described by this method
(and possibly cannot be described by any conjunctive grammar at all).
This center marker stands for
\emph{the middle part of the program
between the declaration of $w$ and the reference to $w$},
and as long as the beginning and the end of this middle part
is distinguishable from $w$,
the same grammar will work.

The last thing to demonstrate
is the use of negation in Boolean grammars.
Consider the following variant of Example~\ref{anb_k_example},
in which one has to ensure
that the first block is \emph{not} equal
to any subsequent block.
This is achieved by putting negation
over the identifier comparison.

\begin{example}\label{anb_k_boolean_example}
The following Boolean grammar describes the language
$\set{a^{n_1} b a^{n_2} b \ldots a^{n_k} b}{k \geqslant 1, \: n_1, \ldots, n_k \geqslant 0, \: n_2 \neq n_1, \ldots, n_k \neq n_1}$.
\begin{equation*}\begin{array}{rcl}
	S &\to& SA \And \lnot Cb \ | \ A \\
	A &\to& aA \ | \ b \\
	C &\to& aCa \ | \ B \\
	B &\to& BA \ | \ b
\end{array}\end{equation*}
\end{example}

This grammar can be rewritten without using negation
by replacing $C$
with a new nonterminal symbol
that defines identifier inequality.

\section{A model programming language and its grammar} \label{section_the_grammar}

For a quick introduction
into the model programming language used in this paper,
consider the following sample program in this language.

\begin{quote}
	\footnotesize
\begin{verbatim}
	average(x, y) { return (x+y)/2; }
	factorial(n)
	{
	   var i, product;
	   i=1;
	   product=1;
	   while(i<=n) {
	      product=product*i;
	      i=i+1;
	   }
	   return product;
	}
	factorial2(n)
	{
	   if(n>=2)
	      return n*factorial2(n-1);
	   else
	      return 1;
	}
	main(arg)
	{
	   return average(factorial(arg), factorial2(arg));
	}
\end{verbatim}
\end{quote}

This is a well-formed program.
All functions and variables
are defined before their use.
The number of arguments in each function call
matches that in the definition of that function:
for example, \texttt{average} is defined with two arguments
and called with two arguments.
Each variable declaration has its scope of visibility:
if \texttt{product} were declared and initialized
inside the \texttt{while} statement,
then the statement \texttt{return product;}
would refer to an undeclared variable.
There are no duplicate declarations.

In the rest of this section,
a semi-formal definition
of the syntax of this language is presented,
with every point immediately expressed
in the formalism of Boolean grammars.

\subsection{Alphabet}\label{section_the_grammar__alphabet}

A program is a finite string over an alphabet $\Sigma$
that consists of the following 54 characters:
	26 letters (\grt{a}, \ldots, \grt{z}),
	10 digits (\grt{0}, \ldots, \grt{9}),
	the space (\grt{\grspace}),
	and 17 punctuators
	(``\grt{(}'', ``\grt{)}'', \grt{\{}, \grt{\}},
	``\grt{,}'', ``\grt{;}'',
	\grt{+}, \grt{-}, \grt{*}, \grt{/}, \grt{\%},
	\grt{\&}, \grt{|}, \grt{!}, \grt{=}, \grt{<}, \grt{>}).
All whitespace characters, such as the newline and the tab,
are treated as space.

The Boolean grammar to be constructed
defines strings over this alphabet $\Sigma$.
The grammar has the following 7 nonterminals
that define some subsets of the alphabet referenced in the grammar.
\begin{align*}
	\anyletter &\to \text{``\grt{a}''} \ | \ \ldots \ | \ \text{``\grt{z}''}
		\\
	\anydigit &\to \text{``\grt{0}''} \ | \ \ldots \ | \ \text{``\grt{9}''}
		\\
	\anyletterdigit &\to \anyletter \ | \ \anydigit
		\\
	\anypunctuator &\to \text{``\grt{(}''} \ | \ \text{``\grt{)}''} \ | \
		\text{``\grt{\{}''} \ | \ \text{``\grt{\}}''} \ | \
		\text{``\grt{,}''} \ | \ \text{``\grt{;}''} \ | \
		\text{``\grt{+}''} \ | \ \text{``\grt{-}''} \ | \ \text{``\grt{*}''} \ | \
		\text{``\grt{/}''} \ | \
		\text{``\grt{\&}''} \ | \ \text{``\grt{|}''} \ | \
		\\ &\hspace*{0.7cm}
		\text{``\grt{!}''} \ | \
		\text{``\grt{=}''} \ | \ \text{``\grt{<}''} \ | \ \text{``\grt{>}''} \ | \ \text{``\grt{\%}''}
			\\
	\anychar &\to \grt{\grspace} \ | \ \anyletter \ | \ \anydigit \ | \ \anypunctuator
		\\
	\anypunctuatorexceptrightpar &\to
		\text{``\grt{(}''} \ | \ \text{``\grt{\{}''} \ | \ \text{``\grt{\}}''} \ | \
		\text{``\grt{,}''} \ | \
		\text{``\grt{;}''} \ | \
		\text{``\grt{+}''} \ | \ \text{``\grt{-}''} \ | \
		\text{``\grt{*}''} \ | \ \text{``\grt{/}''} \ | \ \text{``\grt{\&}''} \ | \
		\text{``\grt{|}''} \ | \ \text{``\grt{!}''} \ | \
		\\ &\hspace*{0.7cm}
		\text{``\grt{=}''} \ | \
		\text{``\grt{<}''} \ | \ \text{``\grt{>}''} \ | \ \text{``\grt{\%}''}
			\\
	\anycharexceptbracespace &\to
		\text{``\grt{(}''} \ | \ \text{``\grt{)}''} \ | \ \text{``\grt{,}''} \ | \
		\text{``\grt{;}''} \ | \
		\text{``\grt{+}''} \ | \ \text{``\grt{-}''} \ | \
		\text{``\grt{*}''} \ | \ \text{``\grt{/}''} \ | \ \text{``\grt{\&}''} \ | \
		\text{``\grt{|}''} \ | \ \text{``\grt{!}''} \ | \ \text{``\grt{=}''} \ | \
		\\ &\hspace*{0.7cm}
		\text{``\grt{<}''} \ | \ \text{``\grt{>}''} \ | \ \text{``\grt{\%}''}
		\ | \ \anyletter \ | \ \anydigit
\end{align*}
There are also the following three nonterminal symbols
for simple regular sets of strings over $\Sigma$.
\begin{align*}
	\anystring &\to \anystring\ \anychar \ | \ \epsilon
		\\
	\anyletterdigits &\to \anyletterdigits\ \anyletterdigit \ | \ \epsilon
		\\
	\safeendingstring &\to \anystring\ \anypunctuator \ | \
		\anystring\ \grt{\grspace} \ | \ \epsilon
\end{align*}
The last nonterminal, $\safeendingstring$, denotes a string that can be directly followed
by an indentifier or a keyword.
It shall be used to ensure
that names are never erroneously split in two,
so that, for instance, an expression \texttt{varnish;}
is never mistaken for a declaration \texttt{var nish;}.

In a usual parsing framework, this kind of conditions
are be ensured by the ``longest match'' principle,
according to which, a longer sequence of symbols
should always be preferred to a shorter one.
However, there is no way to attach this principle to a grammar,
and the correct splitting of a program into tokens
has to be expressed entirely within grammar rules.

\subsection{Lexical conventions}\label{section_the_grammar__lexical}

The character stream is separated into \emph{tokens} from left to right,
with possible whitespace characters between them.
Each time the longest possible sequence of characters
that forms a token is consumed,
or a whitespace character is discarded.
%
%
The handling of whitespace is facilitated
by a nonterminal $\WS$
representing a possibly empty sequence of whitespace characters.
\begin{equation*}
	\WS \to \WS\ \grt{\grspace}\ | \ \epsilon
\end{equation*}

This programming language has 28 token types.
In the grammar, each of them
is represented by a nonterminal symbol enclosed in frame
($\tNotEqual$, etc.),
which defines the set of all valid tokens of this type,
possibly followed by whitespace characters.

First, there are 5 keywords:
\texttt{var}, \texttt{if}, \texttt{else}, \texttt{while}, \texttt{return}.
\begin{align*}
	\Keyword &\to
		\tVar\ \big|\ \tIf\ \big|\ \tElse\ \big|\ \tWhile\ \big|\ \tReturn
			\\
	\tVar &\to
		\grt{v}\ \grt{a}\ \grt{r}\ \WS
			\\
	\tIf &\to
		\grt{i}\ \grt{f}\ \WS
			\\
	\tElse &\to
		\grt{e}\ \grt{l}\ \grt{s}\ \grt{e}\ \WS
			\\
	\tWhile &\to
		\grt{w}\ \grt{h}\ \grt{i}\ \grt{l}\ \grt{e}\ \WS
			\\
	\tReturn &\to
		\grt{r}\ \grt{e}\ \grt{t}\ \grt{u}\ \grt{r}\ \grt{n}\ \WS
\end{align*}

An \emph{identifier}
is a finite nonempty sequence of letters and digits 
that begins with a letter
and is not a keyword.
In a usual parsing framework,
this could be specified by defining a certain order of preference between the rules.
Boolean grammars are powerful enough to express this order explicitly,
leading to the following description
of the set of identifiers
exactly according its definition.
\begin{align*}
	\tId &\to \tIdAux\, \WS \And \lnot \Keyword
		\\
	\tIdAux &\to \anyletter \ | \ \tIdAux\ \anyletter \ | \ \tIdAux\ \anydigit
\end{align*}

A \emph{number} is a finite nonempty sequence of digits.
\begin{align*}
	\tNum &\to \tNumAux\ \WS
		\\
	\tNumAux &\to \tNumAux\ \anydigit \ | \ \anydigit
\end{align*}

There are 21 tokens built from punctuator characters,
namely, 13 binary infix operators
(``\texttt{+}'', ``\texttt{-}'', ``\texttt{*}'', ``\texttt{/}'', ``\texttt{\%}'',
``\texttt{\&}'', ``\texttt{|}'',
``\texttt{<}'', ``\texttt{>}'', ``\texttt{<=}'', ``\texttt{>=}'', ``\texttt{==}'', ``\texttt{!=}),
2 unary prefix operators
(``\texttt{-}'', ``\texttt{!}''),
the assignment operator (``\texttt{=}''),
the comma (``\texttt{,}''),
the semicolon (``\texttt{;}''),
parentheses (``\texttt{(}'', ``\texttt{)}'')
and figure brackets (``\texttt{\{}'', ``\texttt{\}}'').
\begin{equation*}
\begin{array}{r@{}c@{\:}l}
	\tPlus\ &\to& \text{``\grt{+}''}\ \WS \\
	\tMinus\ &\to& \text{``\grt{-}''}\ \WS \\
	\tStar\ &\to& \text{``\grt{*}''}\ \WS \\
	\tSlash\ &\to& \text{``\grt{/}''}\ \WS \\
	\tMod\ &\to& \text{``\grt{\%}''}\ \WS
\end{array}~\;~\begin{array}{r@{}c@{\:}l}
	\tNot\ &\to& \text{``\grt{!}''}\ \WS \\
	\tAnd\ &\to& \text{``\grt{\&}''}\ \WS \\
	\tOr\ &\to& \text{``\grt{|}''}\ \WS \\
	\tLessThan\ &\to& \text{``\grt{<}''}\ \WS \\
	\tGreaterThan\ &\to& \text{``\grt{>}''}\ \WS
\end{array}~\;~\begin{array}{r@{}c@{\:}l}
	\tGreaterEqual\ &\to& \text{``\grt{>}''} \ \text{``\grt{=}''}\ \WS \\
	\tLessEqual\ &\to& \text{``\grt{<}''} \ \text{``\grt{=}''}\ \WS \\
	\tEqual\ &\to& \text{``\grt{=}''} \ \text{``\grt{=}''}\ \WS \\
	\tNotEqual\ &\to& \text{``\grt{!}''} \ \text{``\grt{=}''}\ \WS \\
	\tAssign\ &\to& \text{``\grt{=}''}\ \WS
\end{array}~\;~\begin{array}{r@{}c@{\:}l}
	\tRightPar\ &\to& \text{``\grt{)}''}\ \WS \\
	\tLeftPar\ &\to& \text{``\grt{(}''}\ \WS \\
	\tLeftBrace\ &\to& \text{``\grt{\{}''}\ \WS \\
	\tRightBrace\ &\to& \text{``\grt{\}}''}\ \WS
\end{array}~\;~\begin{array}{r@{}c@{\:}l}
	\tSemicolon\ &\to& \text{``\grt{;}''}\ \WS \\
	\tComma\ &\to& \text{``\grt{,}''}\ \WS
\end{array}
\end{equation*}

\subsection{Identifier matching}\label{section_the_grammar__identifier}

The grammar for the programming language
often has to specify that two identifiers are identical.
The general possibility of doing that
was demonstrated in Example~\ref{wcw_example} above.
A version of that example
adapted to handle identifiers and token separation rules
in the model programming language
shall now be constructed.

The nonterminal symbol $\grC$
defines the set of all strings $wxwy$,
where $w$ is an identifier,
$x$ is an arbitrarily long middle part of the program
between these two identifiers,
and $y$ is a possibly empty sequence of whitespace characters.
\begin{align*}
	\grC &\to \Clen \And \Citerate \ | \ \grC\ \grt{\grspace} \\
	\Clen &\to \anyletterdigit\ \Clen\ \anyletterdigit \ | \
		\anyletterdigit\ \Cmid\ \anyletterdigit
\intertext{%
The form of the middle part $x$ in $wxwy$ is ensured by $\Cmid$,
which verifies that $w$ is the longest prefix
and the longest suffix of $wxw$
that is formed of letters and digits.
}
	\Cmid &\to \grt{\grspace} \ | \ \anypunctuator\ \anystring\ \anypunctuator \ | \
		\grt{\grspace}\ \anystring\ \grt{\grspace}
			\\
	\Cmid &\to 
		\grt{\grspace}\ \anystring\ \anypunctuator \ | \
		\anypunctuator\ \anystring\ \grt{\grspace} \ | \
		\anypunctuator
\end{align*}
In the definition of $\Citerate$,
each nonterminal symbol $C_\sigma$,
with $\sigma \in \{\grt{a}, \ldots, \grt{z}, \grt{0}, \ldots, \grt{9}\}$,
specializes in comparing one particular character
(cf.~$A$ and $B$ in Example~\ref{wcw_example},
which could be called $C_a$ and $C_b$).
\begin{align*}
	\Citerate &\to
		\Cc{$\sigma$}\, \sigma \And \Citerate\ \sigma
		&& (\text{for all } \sigma \in \{\grt{a}, \ldots, \grt{z}, \grt{0}, \ldots, \grt{9}\})
			\\
	\Citerate &\to
		\anyletterdigits\ \Cmid
			\\
	\Cc{$\sigma$} &\to \anyletterdigit\ \Cc{$\sigma$}\ \anyletterdigit 
		&& (\text{for all } \sigma \in \{\grt{a}, \ldots, \grt{z}, \grt{0}, \ldots, \grt{9}\})
			\\
	\Cc{$\sigma$} &\to \sigma\ \anyletterdigits\ \Cmid
		&& (\text{for all } \sigma \in \{\grt{a}, \ldots, \grt{z}, \grt{0}, \ldots, \grt{9}\})
\end{align*}

As the first application of identifier comparison,
consider describing a list of pairwise unique identifiers.
In the sample program
given in the beginning of this section,
there is a function header \texttt{average(x, y)}
and a variable declaration statement \texttt{var i, prod;}.
Each of them contains a list of identifiers being declared,
and no identifier may appear on the list twice.

Such lists are described by a nonterminal symbol $\grsListOfDistinctIds$
in a way that reminds of Example~\ref{anb_k_example},
although this time,
instead of comparing all elements to the first element,
the grammar compares
\emph{every pair} of identifiers on the list,
ensuring that they are all distinct.
This is done by a two-level iteration:
first, $\grsListOfDistinctIds$ sets up a comparison
of every element to all previous elements,
to be carried out by the nonterminal \gra{no-multiple-declarations}.
\begin{align*}
	\grsListOfDistinctIds &\to
		\grsListOfDistinctIds\ \tComma\ \tId \And \gra{no-multiple-declarations}
		\ | \
		\tId
\intertext{%
On the second level of iteration,
\gra{no-multiple-declarations},
applied to a prefix of the list,
ensures that no previous element
coincides with the last element of the prefix.
}
	\gra{no-multiple-declarations} &\to
		\tId\ \tComma\ \gra{no-multiple-declarations} \And \lnot\grC
		\ | \
		\tId
\end{align*}
The actual test for inequality
is done by appropriately negating $\grC$.

\subsection{Expressions}\label{section_the_grammar__expressions}

Arithmetical expressions
are formed of identifiers and constant numbers
using binary operators,
unary operators,
brackets,
function calls
and assignment.
The definition of an expression ($\grsExpr$)
is formalized in the grammar
in the standard way.
No efforts are yet made to make the grammar unambiguous,
and the grammar given here allows the operators to be evaluated in any order.

\textbf{Basic expressions:}
any identifier is an expression,
and any number is an expression.
\begin{equation*}
	\grsExpr \to \tId \ | \ \tNum
\end{equation*}

\textbf{Expression enclosed in brackets:}
$\texttt{(} e \texttt{)}$ is an expression
for every expression $e$.
\begin{equation*}
	\grsExpr \to \tLeftPar\ \grsExpr\ \tRightPar
\end{equation*}

\textbf{Binary operation:}
$e_1 \mathop{op} e_2$ is an expression
for every binary operator $op$
and for all expressions $e_1$, $e_2$.
\begin{align*}
	\grsExpr \to 
		\grsExpr\ op\ \grsExpr
	&& (op \in \{
	\tPlus,
	\tMinus,
	\tStar,
	\tSlash,
	\tMod,
	\tAnd,
	\tOr,
	\tLessThan,
	\tGreaterThan,
	\tGreaterEqual,
	\tLessEqual,
	\tEqual,
	\tNotEqual\})
\end{align*}

\textbf{Unary operation:}
$\mathop{op}e$ is an expression
for every unary operator
and for all expressions $e$.
\begin{equation*}
	\grsExpr \to 
		\tMinus\ \grsExpr \ | \ \tNot\ \grsExpr
\end{equation*}

\textbf{Assignment:}
$x$ \texttt{=} $e$ is an expression
for every identifier $x$
and expression $e$.
\begin{equation*}
	\grsExpr \to 
		\tId\ \tAssign\ \grsExpr
\end{equation*}

\textbf{Function call:}
$f$ \texttt{(} $e_1$ \texttt{,} \ldots \texttt{,} $e_k$ \texttt{)} is an expression
for all identifiers $f$
and expressions $e_1, \ldots, e_k$, with $k \geqslant 0$.
This is a call to the function $f$
with the arguments $e_1, \ldots, e_k$.
For later reference,
it is also denoted by a separate nonterminal symbol $\grsExprCall$.
\begin{align*}
	\grsExpr &\to \grsExprCall
		\\
	\grsExprCall &\to \tId\ \tLeftPar\ \grsListOfExpr\ \tRightPar
\intertext{%
The nonterminal $\grsListOfExpr$ defines (possibly empty) lists of expressions
separated by commas.
}
	\grsListOfExpr &\to \grsListOfExprOnePlus \ | \ \epsilon
		\\
	\grsListOfExprOnePlus &\to \grsListOfExprOnePlus\ \tComma\ \grsExpr \ | \ \grsExpr
\end{align*}

\subsection{Statements}\label{section_the_grammar__statements}

The model programming language
has the following six types of statements ($\grsStatement$).
The rules of the grammar
defining their form
are standard.
The rules describing the conditional statement 
have the \emph{dangling else} ambiguity;
this will be corrected later in Section~\ref{two_standard_ambiguous}.

\textbf{Expression-statement:}
$e$ \texttt{;} is a statement for every expression $e$.	
\begin{align*}
	\grsStatement &\to
		\grsExpr\ \tSemicolon
\intertext{%
\textbf{Compound statement:}
$\texttt{\{} \: s_1 \: s_2 \: \ldots \: s_k \: \texttt{\}}$ is a statement
for all $k \geqslant 0$ and for all statements $s_1, \ldots, s_k$.
}	
	\grsStatement &\to
		\tLeftBrace\ \grsStatements\ \tRightBrace
			\\
	\grsStatements &\to
		\grsStatements\ \grsStatement\ \big|\ \epsilon
\intertext{%
\textbf{Conditional statement:}
\texttt{if} \texttt{(} $e$ \texttt{)} $s$
and
\texttt{if} \texttt{(} $e$ \texttt{)} $s$ \texttt{else} $s'$
are statements for every expression $e$
and for all statements $s$, $s'$.
}
	\grsStatement &\to
		\tIf\ \tLeftPar\ \grsExpr\ \tRightPar\ \grsStatement
			\\
	\grsStatement &\to
		\tIf\ \tLeftPar\ \grsExpr\ \tRightPar\ \grsStatement\ \tElse\ \grsStatement
\intertext{%
\textbf{Iteration statement:}
\texttt{while} \texttt{(} $e$ \texttt{)} $s$
is a statement for every expression $e$
and statement $s$.
}
	\grsStatement &\to
		\tWhile\ \tLeftPar\ \grsExpr\ \tRightPar\ \grsStatement
\intertext{%
\textbf{Declaration statement:}
$\texttt{var} \; x_1\texttt{,} \ldots \texttt{,}x_k\texttt{;}$ is a statement
for every $k \geqslant 1$ and for all identifiers $x_1, \ldots, x_k$.
Declaration statements are also
denoted by a separate nonterminal symbol $\grsStatementVar$.
}
	\grsStatement &\to
		\grsStatementVar
			\\
	\grsStatementVar &\to
		\tVar\ \grt{\grspace}\ \grsListOfDistinctIds\ \tSemicolon
\intertext{%
\textbf{Return statement:}
\texttt{return} $e$ \texttt{;}
is a statement for every expression $e$.
}	
	\grsStatement &\to
		\grsStatementReturn
			\\
	\grsStatementReturn &\to
		\tReturn\ \grsExpr\ \tSemicolon
		\And
		\returnstatementfix
			\\
	\returnstatementfix &\to
		\grt{r}\ \grt{e}\ \grt{t}\ \grt{u}\ \grt{r}\ \grt{n}\ \anypunctuator\ \anystring
		\ | \
		\grt{r}\ \grt{e}\ \grt{t}\ \grt{u}\ \grt{r}\ \grt{n}\ \grt{\grspace}\ \anystring
\end{align*}
A conjunction with $\returnstatementfix$
is necessary to ensure that the keyword \texttt{return}
is always followed by some punctuator or whitespace character.
Only in this case, a string may be parsed as a return statement.
Without this condition,
the first conjunct
	$\tReturn\ \grsExpr\ \tSemicolon$
would define, for instance, a string \texttt{returnable;},
as if it were a return statement \texttt{return able;}.
This detail could be managed without using conjunction,
but is a more complicated way.

The grammar also defines
a subclass of \emph{returning statements} ($\grsStatementRet$),
that is, those that may terminate their execution
only by a return statement.
A return statement itself is returning.
\begin{align*}
	\grsStatementRet &\to
		\grsStatementReturn
\intertext{%
A conditional statement is returning,
if both branches are returning.
}
	\grsStatementRet &\to
		\tIf\ \tLeftPar\ \grsExpr\ \tRightPar\ \grsStatementRet\ \tElse\ \grsStatementRet
\intertext{%
A compound statement is returning,
if so is its last constituent.
The set of returning compound statements
is also denoted by a separate nonterminal symbol,
$\grsStatementCompoundRet$.
}
	\grsStatementRet &\to
		\grsStatementCompoundRet
			\\
	\grsStatementCompoundRet &\to
		\tLeftBrace\ \grsStatements\ \grsStatementRet\ \tRightBrace
\end{align*}

\subsection{Function declarations}\label{section_the_grammar__functions}

A function declaration
begins with a \emph{header}
of the form $f$ \texttt{(} $x_1\texttt{,} \ldots \texttt{,}x_k$ \texttt{)},
where $k \geqslant 0$
and $f, x_1, \ldots, x_k$ are identifiers.
The identifier $f$ is the \emph{name} of the function,
and the identifiers $x_1, \ldots, x_k$ are its \emph{formal arguments}.
The arguments must be distinct,
hence the grammar refers to a list of pairwise distinct identifiers
($\grsListOfDistinctIds$).
\begin{align*}
	\grsFunctionHeader &\to
		\tId\ \tLeftPar\ \grsListOfDistinctIds\ \tRightPar
			\\
	\grsFunctionHeader &\to
		\tId\ \tLeftPar\ \tRightPar
\intertext{%
A \emph{function declaration} ($\grsFunction$)
is a header ($\grsFunctionHeader$)
followed by a returning compound statement ($\grsStatementCompoundRet$),
called the \emph{body} of the function.
}
	\grsFunction &\to
		\grsFunctionHeader\ \grsStatementCompoundRet
		\And
		\tId\ \tLeftPar\ \gra{all-variables-declared}
\end{align*}
The second conjunct in the latter rule
refers to the nonterminal symbol $\gra{all-variables-declared}$
representing the conditions on variable declaration.
Intuitively, one can interpret this rule
in the sense that $\grsFunctionHeader\ \grsStatementCompoundRet$
first defines a certain structure,
and then $\gra{all-variables-declared}$
processes that structure
to verify declaration of variables before use.
Then, for every string being defined by $\gra{all-variables-declared}$,
one can assume that it is already of the form $\grsFunctionHeader\ \grsStatementCompoundRet$.
This makes the rules for $\gra{all-variables-declared}$,
presented below,
easier to construct and understand.

\subsection{Declaration of variables before use}\label{section_the_grammar__variable_declaration}

For each function, the goal is to check
that every reference to a variable in the function body
is preceded by a declaration of a variable with the same name.
A \emph{reference} is an identifier occurring in an expression.
\emph{Declarations} take place in the list of function arguments
and in \texttt{var} statements;
in the latter case,
the reference should be \emph{in the scope} of this declaration,
that is, within the same compound statement
as the \texttt{var} statement
or in any nested statements,
and occurring later than the \texttt{var} statement.
Another related thing to check
is that no declaration is in the scope of another declaration
of the same variable.
The purpose of the nonterminal symbol $\gra{all-variables-declared}$
is to check all these conditions
for a particular function.

\begin{figure}[t]
	\centerline{\includegraphics[scale=0.8]{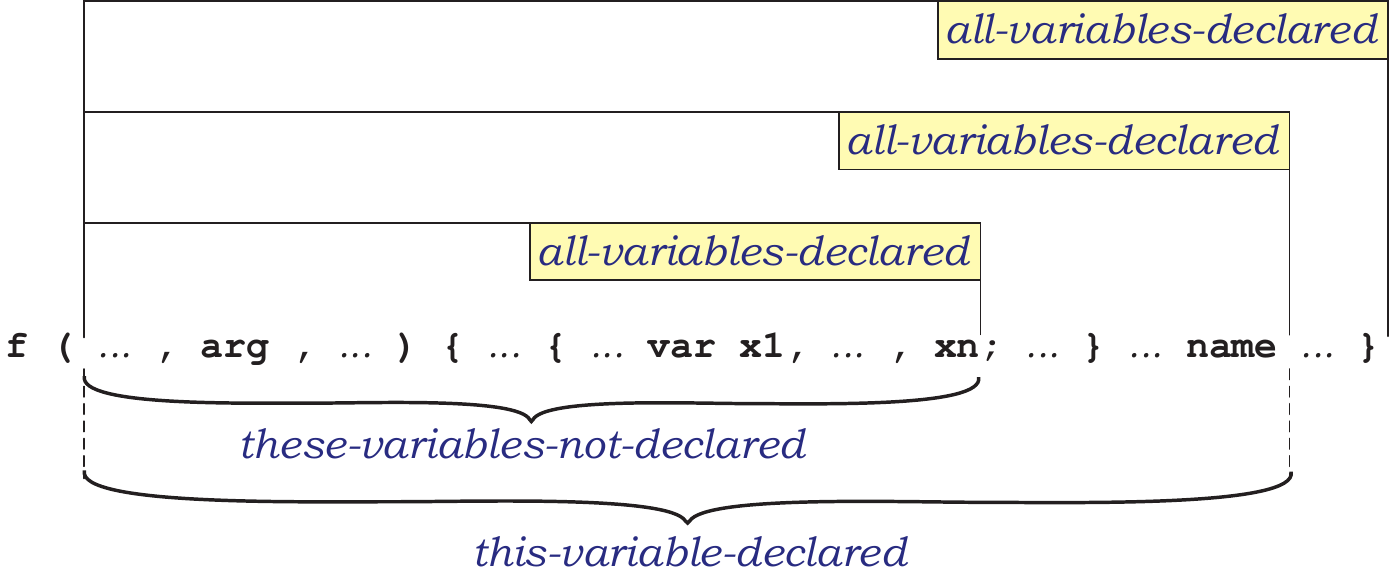}}
	\caption{How \gra{all-variables-declared} processes all prefixes of the function body,
		applying \gra{these-variables-not-declared} to declarations
		and \gra{this-variable-declared} to references.}
	\label{f:all_variables_declared}
\end{figure}

The rules for the nonterminal $\gra{all-variables-declared}$
iterate over \emph{all prefixes (of the function body) that end with an identifier},
as illustrated in Figure~\ref{f:all_variables_declared}.
This is done generally in the same way as in Example~\ref{anb_k_example},
with the following details to note.
First, the function body is split into tokens again,
and all irrelevant tokens are skipped.
Special care has to be exercised
when skipping a number or a keyword,
because the characters forming them
might actually be a suffix of an identifier to be checked;
this possibility is ruled out in the rule for \gra{all-variables-declared-safe}.
\begin{align*}
	\gra{all-variables-declared} &\to
		\gra{all-variables-declared-safe}\ \tNum
			\\
	\gra{all-variables-declared} &\to
		\gra{all-variables-declared-safe}\ \Keyword
			\\
	\gra{all-variables-declared} &\to
		\gra{all-variables-declared-safe}\ \tId\ \tLeftPar
			\\
	\gra{all-variables-declared-safe} &\to
		\gra{all-variables-declared}
		\And
		\safeendingstring
\intertext{%
Any punctuator character is skipped,
unless it is a semicolon
concluding a \texttt{var} statement.
}
	\gra{all-variables-declared} &\to
		\gra{all-variables-declared}\ \anypunctuator\ \WS
		\And
		\lnot \safeendingstring\ \grsStatementVar
\intertext{%
It is important to distinguish
between identifiers representing declarations
and identifiers representing references.
Once a \texttt{var} statement is found,
$\gra{these-variables-not-declared}$ shall verify
that none of the variables defined here
have previously been defined;
this is one of the two cases illustrated in Figure~\ref{f:all_variables_declared}.
}
	\gra{all-variables-declared} &\to
		\gra{these-variables-not-declared}\ \tSemicolon
		\And
		\gra{all-variables-declared-safe}\ \grsStatementVar
\intertext{%
The other case shown in Figure~\ref{f:all_variables_declared}
is that for each reference found,
a nonterminal $\gra{this-variable-declared}$ is invoked
to check that this variable has an earlier declaration.
}
	\gra{all-variables-declared} &\to
		\gra{this-variable-declared}
		\And
		\gra{all-variables-declared-safe}\ \tId
\intertext{%
Finally, once the whole function body is processed,
only the list of arguments in its header remains.
This list may be empty,
hence there are two terminating rules.
}
	\gra{all-variables-declared} &\to
		\grsListOfDistinctIds\ \tRightPar
		\ \big|\
		\tRightPar
\end{align*}

\begin{figure}[t]
	\centerline{\includegraphics[scale=0.8]{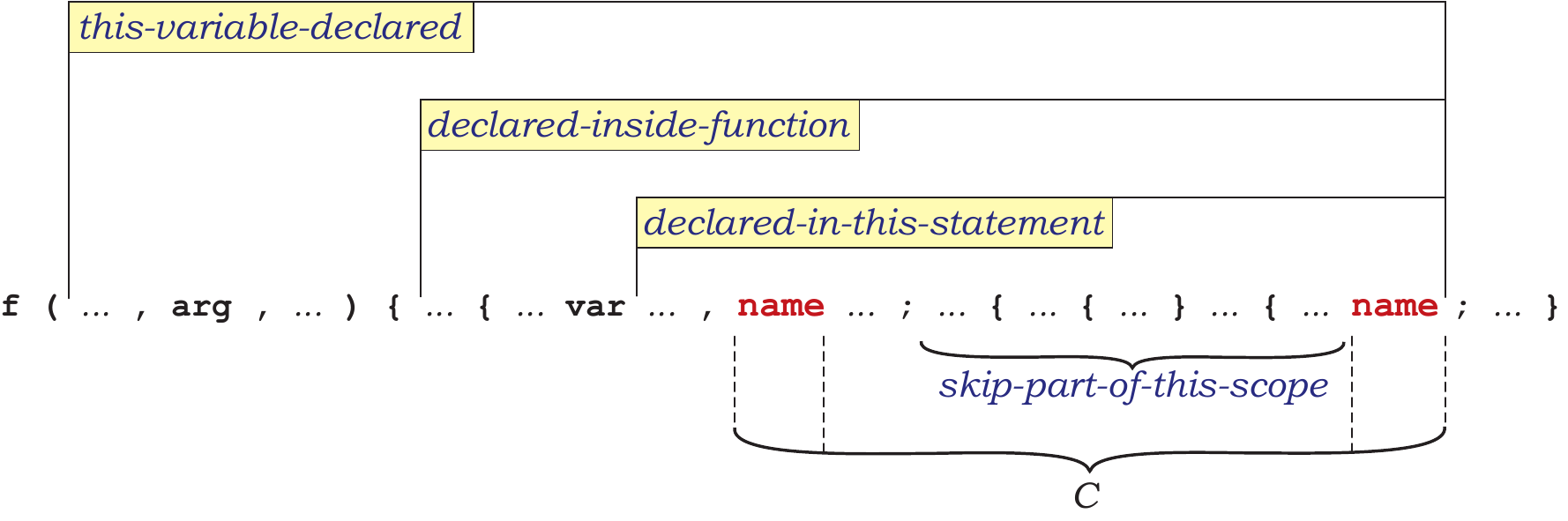}}
	\caption{How \gra{this-variable-declared}, \gra{declared-inside-function}
		and \gra{declared-in-thjs-statement}
		look for a matching declaration.}
	\label{f:this_variable_declared}
\end{figure}

Thus, for every prefix ending with a reference to a variable,
the nonterminal \gra{this-variable-declared} is used to match it
to a declaration of a variable that occurs inside this prefix.
This process is illustrated in Figure~\ref{f:this_variable_declared}.
This time, the rules iterate over \emph{all suffixes of the current prefix},
beginning at different tokens
and ending with the identifier being checked.
First, the rules for \gra{this-variable-declared}
search for a declaration among the function's arguments,
and if it is found,
it is left to match the identifiers using $\grC$.
\begin{align*}
	\gra{this-variable-declared} &\to
		\tId\ \tComma\ \gra{this-variable-declared}
		\ \big|\
		\grC
\intertext{%
If failed,
the nonterminal \gra{declared-inside-function} is invoked
to look for a declaration in a suitable \texttt{var} statement.
}
	\gra{this-variable-declared} &\to
		\tId\ \tRightPar\ \tLeftBrace\ \gra{declared-inside-function}
\intertext{%
If the function has no arguments,
the search for a \texttt{var} statement in the body
begins by the following rule.
}
	\gra{this-variable-declared} &\to
		\tRightPar\ \tLeftBrace\ \gra{declared-inside-function}
\end{align*}

While searching for a \texttt{var} statement,
variable scopes have to be observed,
and for that purpose, the rules for \gra{declared-inside-function}
parse the function body
according to the nested structure of statements.
First, any complete statement may be ignored:
this means that the desired variable is not declared there.
\begin{align*}
	\gra{declared-inside-function} &\to
		\grsStatement\ \gra{declared-inside-function}
\intertext{%
If the reference being checked
is inside a compound statement,
then the following rule
moves the search one level deeper
into a nested compound statement.
}
	\gra{declared-inside-function} &\to
		\tLeftBrace\ \gra{declared-inside-function}
\intertext{%
For \texttt{if} and \texttt{while} statements,
a nested scope is entered
through an extra nonterminal \gra{declared-inside-function-nested},
which indicates that the current statement
is not directly within a compound statement.
}
	\gra{declared-inside-function} &\to
		\tIf\ \tLeftPar\ \grsExpr\ \tRightPar\
			\gra{declared-inside-function-nested}
			\\
	\gra{declared-inside-function} &\to
		\tIf\ \tLeftPar\ \grsExpr\ \tRightPar\ \grsStatement\
			\tElse\ \gra{declared-inside-function-nested}
			\\
	\gra{declared-inside-function} &\to
		\tWhile\ \tLeftPar\ \grsExpr\ \tRightPar\
			\gra{declared-inside-function-nested}
\intertext{%
The rules for \gra{declared-inside-function-nested}
process potentially nested \texttt{if} and \texttt{while} statements
and get back to \gra{declared-inside-function}
as soon as a compound statement begins.
}
	\gra{declared-inside-function-nested} &\to
		\tLeftBrace\ \gra{declared-inside-function}
			\\
	\gra{declared-inside-function-nested} &\to
		\tIf\ \tLeftPar\ \grsExpr\ \tRightPar\
			\gra{declared-inside-function-nested}
			\\
	\gra{declared-inside-function-nested} &\to
		\tIf\ \tLeftPar\ \grsExpr\ \tRightPar\ \grsStatement\
			\tElse\ \gra{declared-inside-function-nested}
			\\
	\gra{declared-inside-function-nested} &\to
		\tWhile\ \tLeftPar\ \grsExpr\ \tRightPar\
			\gra{declared-inside-function-nested}
\end{align*}

If a \texttt{var} statement is encountered,
the desired declaration may be there.
In this case, the nonterminal \gra{declared-inside-function}
is used to find the correct declaration
among the variables listed in this \texttt{var} statement,
and $\grC$ is invoked to match identifiers.
\begin{align*}
	\gra{declared-inside-function} &\to
		\tVar\ \grt{\grspace}\ \gra{declared-in-this-statement}
			\\
	\gra{declared-in-this-statement} &\to
		\tId\ \tComma\ \gra{declared-in-this-statement}
			\\
	\gra{declared-in-this-statement} &\to
		\grC
		\And
		\gra{ignore-remaining-variables}\ \gra{skip-part-of-this-scope}
\intertext{%
After skipping the remaining variables
declared in this \texttt{var} statement
(\gra{ignore-remaining-variables}),
the middle part between the declaration and the reference
is described by a nonterminal \gra{skip-part-of-this-scope},
which ensures that the reference
stays in the scope of the declaration.
}
	\gra{ignore-remaining-variables} &\to
		\tId\ \tComma\ \gra{ignore-remaining-variables} \ | \
		\tId\ \tSemicolon
			\\
	\gra{skip-part-of-this-scope} &\to
		\gra{skip-part-of-this-scope}\ \tLeftBrace\ \grsStatements\ \tRightBrace
			\\
	\gra{skip-part-of-this-scope} &\to
		\gra{skip-part-of-this-scope}\ \tLeftBrace
			\\
	\gra{skip-part-of-this-scope} &\to
		\gra{skip-part-of-this-scope}\ \anycharexceptbracespace\ \WS
\end{align*}

Now consider the other nonterminal \gra{these-variables-not-declared},
which is used in the rules for \gra{all-variables-declared}
for any prefix ending with a declaration,
in order to ensure that
none of these variables have been declared before.
Here negation comes in particularly useful,
because the condition that an identifier
is in the scope of a variable
with the same name
has already been expressed as \gra{this-variable-declared},
and now it is sufficient to negate it.
The rules for \gra{these-variables-not-declared}
iterate over all variables declared at this point.
\begin{align*}
	\gra{these-variables-not-declared} &\to
		\gra{these-variables-not-declared}\ \tComma\ \tId\
		\And
		\lnot\gra{this-variable-declared}
\intertext{%
The iteration terminates
after checking the first variable declared in this \texttt{var} statement.
}
	\gra{these-variables-not-declared} &\to
		\safeendingstring\ \tVar\ \grt{\grspace}\ \tId
		\And
		\lnot\gra{this-variable-declared}
\end{align*}

\subsection{Declaration of functions before use}\label{section_the_grammar__function_declaration}

Another kind of references
to be matched to their declarations
are calls to functions.
Whenever a function is called,
somewhere earlier in the program
there should be a function header
that opens a declaration of a function
with the same name and with the same number of arguments.
Furthermore, a program may not contain multiple declarations of functions
sharing the same name and the same number of arguments.

Checking these conditions
requires matching each function call
to a suitable earlier function declaration.
Similarly to the rules for \gra{all-variables-declared},
this is done
by considering all prefixes of the program
that end with a function call or a function header.
For that purpose,
all tokens except right parentheses
are being skipped.
\begin{align*}
	\gra{function-declarations} &\to
		\gra{function-declarations}\ \anypunctuatorexceptrightpar\ \WS
			\\
	\gra{function-declarations} &\to
		\gra{function-declarations-safe}\ \Keyword
			\\
	\gra{function-declarations} &\to
		\gra{function-declarations-safe}\ \tId
			\\
	\gra{function-declarations} &\to
		\gra{function-declarations-safe}\ \tNum
			\\
	\gra{function-declarations} &\to
		\epsilon
			\\
	\gra{function-declarations-safe} &\to
		\gra{function-declarations}
		\And
		\safeendingstring
\end{align*}
Whenever a right parenthesis is found,
there are three possibilities,
each handled in a separate rule for \gra{function-declarations}.

\begin{figure}[t]
	\centerline{\includegraphics[scale=0.8]{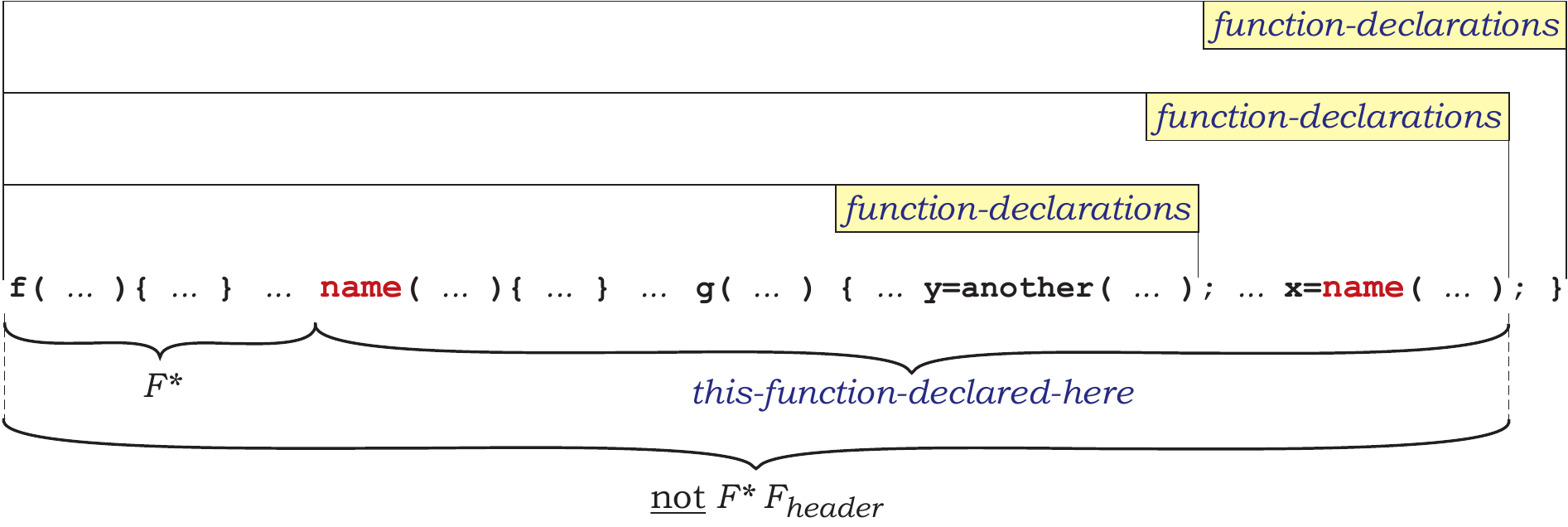}}
	\caption{How \gra{function-declarations} handles a call to a function \texttt{name()}.}
	\label{f:function_declarations__call}
\end{figure}

First, this right parenthesis
could be the last character of a function call expression,
for which one should find a matching function declaration.
This case is identified by the following two conditions:
the current substring should end
with a function call expression ($\grsExprCall$),
and at the same time
the entire substring should not be of the form
$\grsFunctions$ $\grsFunctionHeader$.
The latter condition is essential,
because otherwise (in this model programming language)
a function call is indistinguishable
from a list of arguments in a function header.
\begin{align*}
	\gra{function-declarations} &\to
		\gra{function-declarations}\ \tRightPar
		\And
		\safeendingstring\ \grsExprCall
		\And \\ &\hspace*{5mm}\And
		\lnot \grsFunctions\ \grsFunctionHeader
		\And
		\grsFunctions\ \gra{this-function-declared-here}
\end{align*}
This case is illustrated in Figure~\ref{f:function_declarations__call}.

The concatenation in the last conjunct of the rule
splits the current prefix of the program
into zero or more irrelevant function declarations ($\grsFunctions$)
followed by a substring
that begins with a header of the desired function
and ends with the function call expression,
with these two sharing the same name
and having same number of arguments (\gra{this-function-declared-here}).
The comparison of identifiers (\gra{same-function-name})
and of the number of arguments (\gra{same-number-of-arguments})
is carried out in the following rules.
\begin{align*}
	\gra{this-function-declared-here} &\to
		\gra{same-function-name}
		\And
		\gra{same-number-of-arguments}
			\\
	\gra{same-function-name} &\to
		\grC\ \tLeftPar\ \grsListOfExpr\ \tRightPar
			\\
	\gra{same-number-of-arguments} &\to
		\tId\ \tLeftPar\ \gra{n-of-arg-equal}\ \tRightPar
			\\
	\gra{same-number-of-arguments} &\to
		\tId\ \tLeftPar\ \gra{n-of-arg-equal-0}\ \tRightPar
\intertext{%
Here, the nonterminal \gra{n-of-arg-equal} handles the case of one or more arguments,
whereas \gra{n-of-arg-equal-0} corresponds to the case of a call to a function
with zero arguments.
}
	\gra{n-of-arg-equal-0} &\to
		\tRightPar\ \anystring\ \tLeftPar
			\\
	\gra{n-of-arg-equal} &\to
		\tId\ \tComma\ \gra{n-of-arg-equal}\ \tComma\ \grsExpr
			\\
	\gra{n-of-arg-equal} &\to
		\tId\ \tRightPar\ \anystring\ \tLeftPar\ E
\end{align*}

\begin{figure}[t]
	\centerline{\includegraphics[scale=0.8]{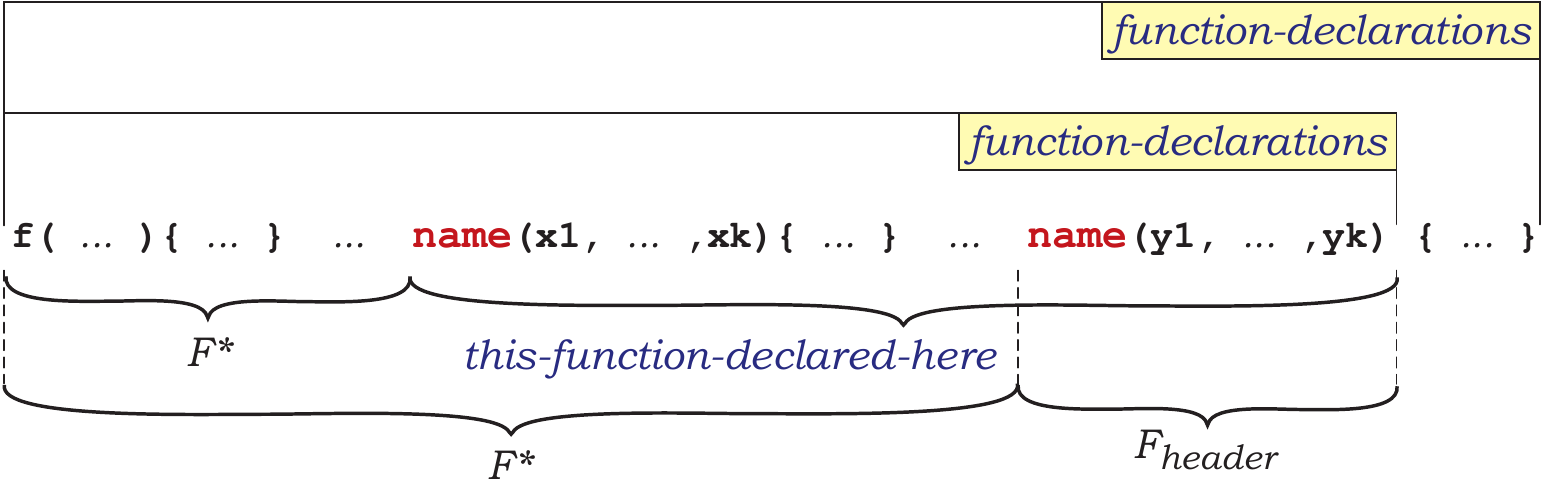}}
	\caption{How \gra{function-declarations} detects duplicate declarations.}
	\label{f:function_declarations__decl}
\end{figure}

The second possibility
with a right parenthesis
encountered in \gra{function-declarations}
is when it is the last character of a function header.
In this case, the grammar should ensure
that no other functions with the same name are declared.
The second conjunct of the following rule
verifies that the current substring ends with a function header
rather than with a function call,
whereas the third conjunct
negates the condition
of having an earlier declaration of the same function.
\begin{equation*}
	\gra{function-declarations} \to
		\gra{function-declarations}\ \tRightPar\
		\And
		\grsFunctions\, \grsFunctionHeader
		\And
		\lnot \grsFunctions\, \gra{this-function-declared-here}
\end{equation*}
Figure~\ref{f:function_declarations__decl} demonstrates
how this rule detects multiple declarations of the same function,
so that the substring is not defined by $\gra{function-declarations}$.
For it to be defined,
it should have no partition
into $\grsFunctions\ \gra{this-function-declared-here}$.

The last case in \gra{function-declarations}
is when a substring ends with a right parenthesis,
but it neither marks an end of a function call expression,
nor is a part of a function header.
This can happen in several ways:
it could be a subexpression enclosed in parentheses,
or a part of a \texttt{for} or a \texttt{while} statement.
In each case,
there is nothing to check,
and the right parenthesis is skipped like any other token.
The case is identified by not ending with $\grsExprCall$.
\begin{equation*}
	\gra{function-declarations} \to
		\gra{function-declarations}\ \tRightPar\
		\And
		\lnot \safeendingstring\, \grsExprCall
\end{equation*}

The check for function declaration before use,
as implemented in the nonterminal symbol $\gra{function-declarations}$,
ensures that both the name and the number of arguments match.
When the same condition is used to define a duplicate declaration,
two declaration are considered duplicate
if they agree both the name in the number of arguments.
This means that functions can be overloaded.

\subsection{Programs}\label{section_the_grammar__programs}

It remains to give the rules
describing the set of well-formed programs
in the model programming language.
A \emph{program} is a finite sequence of function declarations,
which contains one function with the name \texttt{main},
with one argument.

A sequence of function declarations
and a declaration of the main function
are defined by the following rules.
\begin{align*}
	\grsFunctions &\to
		\grsFunctions\ \grsFunction \ | \ \epsilon
			\\
	\grsFunctionMain &\to
		\grt{m}\ \grt{a}\ \grt{i}\ \grt{n}\ \WS\ \tLeftPar\ \tId\ \tRightPar\ \grsStatementCompoundRet
		\And
		\tId\ \tLeftPar\, \gra{all-variables-declared}
\intertext{%
Finally, a single rule for the initial symbol $\grsProgram$
defines what a well-formed program is.
This rule also defines possible whitespace characters
occurring before the first token.
}
	\grsProgram &\to
		\WS\ \grsFunctions\ \grsFunctionMain\ \grsFunctions
		\And
		\WS\ \gra{function-declarations}
\end{align*}

This completes the grammar.

\begin{proposition}
The set of well-formed programs
in the model programming language
is described by a Boolean grammar
with 117 nonterminal symbols
and 361 rules.
\end{proposition}

The grammar constructed above
can be used with any of the several known parsing algorithms
for Boolean grammars.
First, there is a simple extension
of the Cocke--Kasami--Younger algorithm,
with the running time $O(n^3)$
in the length of the input~\cite{BooleanGrammars}.
Like in the case of ordinary grammars,
this algorithm can be accelerated
to run in time $O(n^\omega)$~\cite{BooleanMatrix},
where $\omega<3$ is the exponent
in the complexity of matrix multiplication.
The most practical algorithm
is the GLR~\cite{BooleanLR},
which has worst-case running time $O(n^4)$,
but may run faster for some grammars and inputs,
if a particular parse goes on partially deterministically.

The grammar has been tested
on a large set of positive and negative examples
using one of the existing implementations
of GLR parsing for Boolean grammars~\cite{WhaleCalf}.
The parser contains 754 states
and operates in generally the same way
as GLR parsers for ordinary grammars.

\section{Eliminating negation and ambiguity}\label{section_unambconj}

The grammar for the model programming language
given in Section~\ref{section_the_grammar}
uses negation several times
and contains quite a lot of syntactical ambiguity.
Disregarding these shortcomings
made grammar construction easier.

In general,
unintended ambiguity 
is always undesirable,
and the use of negation can also be viewed as an unnecessary complication.
The purpose of this section
is to explain how the grammar
can be rewritten as an unambiguous conjunctive grammar
describing exactly the same language.
Besides being a conceptually clearer model,
unambiguous conjunctive grammars
also have a better upper bound on the parsing complexity.

\subsection{Negation in auxiliary definitions}

At two occasions,
the grammar uses the negation
in the definitions of basic constructs.
The same definitions now have to be reformulated
without the negation.

First, in Section~\ref{section_the_grammar__lexical},
the rule defining identifiers ($\tId$)
uses negation to describe a regular language
$\{\texttt{a}, \ldots, \texttt{z}\} \{\texttt{a}, \ldots, \texttt{z}, \texttt{0}, \ldots, \texttt{9}\}^* \setminus \{\texttt{var}, \texttt{if}, \texttt{else}, \texttt{while}, \texttt{return}\}$.
The same language can be recognized by a 21-state finite automaton,
which can in turn be simulated in the grammar.

A more interesting use of negation
is in the rule for \gra{no-multiple-declarations},
where it is applied to $\grC$
in order to express identifier inequality.
To eliminate the negation here,
one should define a new nonterminal symbol $\grCneg$
that would describe all strings $uxvy$,
where $u$ and $v$ are distinct identifiers,
$x$ is the middle part of the program between these two identifiers,
and $y$ is a possibly empty sequence of whitespace characters.
The first possibility for $u$ and $v$ not to be equal
is if they are of different length;
this case is handled in the rules for $\Clengt$ ($|u|>|v|$)
and for $\Clenlt$ ($|u|<|v|$).
\begin{align*}
	\grCneg &\to
		\Clenlt\ | \ \Clengt \ | \ \grCneg\ \grt{\grspace} \\
	\Clengt &\to
		\anyletterdigit\ \Clengt \ | \ \anyletterdigit\ \Clen \\
	\Clenlt &\to
		\Clenlt\ \anyletterdigit \ | \ \Clen\ \anyletterdigit
\end{align*}
Otherwise, if $|u|=|v|$,
then $\Citerateneg$ begins comparing the characters of $u$ and $v$
in the same way as done in $\Citerate$,
using $C_\sigma$ to check each character.
\begin{align*}
	\grCneg &\to
		\Clen \And \Citerateneg \\
	\Citerateneg &\to
		\Cc{$\sigma$} \sigma \And \Citerateneg \sigma
			&& (\text{for all } \sigma \in \{\grt{a}, \ldots, \grt{z}, \grt{0}, \ldots, \grt{9}\})
\intertext{
The iteration in $\Citerateneg$ is stopped 
when a pair of mismatched characters is encountered,
that is, if $u=u' \sigma w$ and $v=v' \tau w$,
for some $u'$ and $v'$.
}
	\Citerateneg &\to
		\Cc{$\sigma$} \tau
			&& (\text{for all } \sigma, \tau \in \{\grt{a}, \ldots, \grt{z}, \grt{0}, \ldots, \grt{9}\}, \text{ with } \sigma \neq \tau)
\end{align*}

\subsection{Two standard ambiguous constructs}\label{two_standard_ambiguous}

The rules in Section~\ref{section_the_grammar__expressions}
include two standard cases
of ambiguous definitions in programming languages.
The first of them concerns the expressions.
The given definition is ambiguous,
because the precedence and the associativity of operators
are not defined.
One could make the rules for $\grsExpr$ unambiguous
by rewriting them in the standard way,
introducing a new nonterminal symbol
for each level of precedence.

The rules defining the conditional statement
feature another classical kind of ambiguity,
known as the ``dangling else'' ambiguity.
Indeed, a string such as \texttt{if(x) if(y) s; else t;}
can be parsed in two different ways,
depending on whether the last \texttt{else} clause
binds to the first or to the second \texttt{if} statement.
This ambiguity can also be resolved in the standard way,
by introducing a variant of $\grsStatement$
called $\grsStatementNotIfThen$,
which should define all statements except those of the form if-then,
without an \texttt{else} clause.
Then, all rules for $\grsStatement$
describing statements other than conditional statements
are preserved,
whereas the rules describing conditional statements
take the following form.
\begin{align*}
	\grsStatement &\to
		\tIf\ \tLeftPar\ \grsExpr\ \tRightPar\ \grsStatement
			\\
	\grsStatement &\to
		\tIf\ \tLeftPar\ \grsExpr\ \tRightPar\ \grsStatementNotIfThen\ \tElse\ \grsStatement
\end{align*}
The new nonterminal symbol $\grsStatementNotIfThen$
does not have a rule of the former type (if-then),
but otherwise, it has all the same rules as $\grsStatement$.

\subsection{Variable declarations}\label{section_unambconj__variable_declaration}

The rules requiring declaration of variables before reference,
as given in Section~\ref{section_the_grammar__variable_declaration},
essentially use negation
to ensure that a variable has not been declared before
($\lnot$\gra{this-variable-declared}).
Furthermore, there is some subtle ambiguity
in the rules for \gra{this-variable-declared},
\gra{declared-inside-function}
and \gra{declared-in-this-statement}.
Both shortcomings shall now be corrected
by reimplementing parts of the grammar.

First, consider the ambiguity,
which manifests itself on any program
containing a variable with multiple declarations.
Even though any such program
is ultimately considered ill-formed,
according to Definition~\ref{boolean_unambiguity_definition},
this is still ambiguity,
in the sense that some substrings still have multiple parses.
This affects the complexity of parsing.
Consider the following ill-formed function declaration.
\begin{equation*}
	\texttt{f(}%
	\underbrace{\texttt{arg, arg) \{ var arg; var arg, arg; arg}}_{\gra{this-variable-declared}}%
	\texttt{=0; \}}
\end{equation*}
When checking the reference to \texttt{arg},
the nonterminal \gra{this-variable-declared}
has to handle the underlined substring that ends with that reference.
First, there is an ambiguity between the rules
	\gra{this-variable-declared} $\to$ $\grC$
and 
	\gra{this-variable-declared} $\to$
		\tId\ \tComma\ \gra{this-variable-declared}.
The former matches the first argument of the function
to the reference,
whereas the latter ignores the first argument
and looks for a declaration of \texttt{arg} later.
For this particular string,
both conditions hold at the same time,
hence the ambiguity.
Next, there is a similar ambiguity
between matching the last argument of the function
	(\gra{this-variable-declared} $\to$ $\grC$)
and looking for a declaration of \texttt{arg}
inside the function body
	(\gra{this-variable-declared} $\to$
		\tId\ \tRightPar\ \tLeftBrace\ \gra{declared-inside-function}).
Later on in the grammar, there is an ambiguity
between the first and the second \texttt{var} statements:
in other words, the ambiguity in the choice
between \gra{declared-inside-function} $\to$
		$\grsStatement$ \gra{declared-inside-function}
and
	\gra{declared-inside-function} $\to$
		\tVar\ \grt{\grspace}\ \gra{declared-in-this-statement}.
Finally, when the second \texttt{var} statement
is analyzed by \gra{declared-in-this-statement},
one can use the first or the second identifier in it:
this is the ambiguity
between the two rules for \gra{declared-in-this-statement}.

In each case, the ambiguity could be resolved
by using negation to set the precedence
of the two conditions explicitly.
In general,
given two rules $A \to \alpha$ and $A \to \beta$,
and assuming that the former has higher precedence,
the latter rule would be replaced
with $A \to \beta \And \lnot \alpha$%
~\cite[Prop.~2]{BooleanUnambiguous}.
The goal is to do the same without using negation.

The solution proposed here
is to introduce a negative counterpart
to each of the three nonterminals
that need to be negated
(\gra{this-variable-declared},
\gra{declared-inside-function},
\gra{declared-in-this-statement}).
The rules for this negative counterpart
implement a dual version of the rules
for its original positive version,
and define the opposite condition.
A general transformation
that achieves this effect
is known only for the special class
of \emph{linear conjunctive grammars}~\cite[Thm.~5]{LinearClosure}.
Although the grammar considered here is not exactly linear,
the idea of that general method
shall be used to dualize
the particular rules of this grammar.

The new nonterminal symbols
representing negations of the required conditions
shall be called
\gra{this-variable-not-declared},
\gra{not-declared-inside-function}
and
\gra{not-declared-in-this-statement}.
First, consider how these nonterminals
shall be used in the existing grammar.
The rule for \gra{these-variables-not-declared}
with an explicit negation 
is rewritten using the negative nonterminal.
\begin{equation*}
	\gra{these-variables-not-declared} \to
		\gra{these-variables-not-declared}\ \tComma\ \tId\
		\And
		\gra{this-variable-not-declared}
\end{equation*}
The ambiguity between the first three rules for \gra{this-variable-declared}
is resolved by allowing earlier declarations ($\grC$)
only if there are no later declarations
(\tId\ \tComma\ \gra{this-variable-declared}
or
\tId\ \tRightPar\ \tLeftBrace\ \gra{declared-inside-function}).
Thus, the rule $\gra{this-variable-declared} \to \grC$
is replaced with the following two rules.
\begin{align*}
	\gra{this-variable-declared} &\to
		\grC
		\And
		\tId\ \tComma\ \gra{this-variable-not-declared}
			\\
	\gra{this-variable-declared} &\to
		\grC
		\And
		\tId\ \tRightPar\ \tLeftBrace\ \gra{not-declared-inside-function}
\intertext{%
This prioritizes later declarations over earlier declarations.
The same principle is followed
in disambiguating the definitions of \gra{declared-inside-function}
and \gra{declared-in-this-statement}.
For the former,
every \texttt{var} statement is processed for declarations
only if this variable is not declared later.
}
	\gra{declared-inside-function} &\to
		\tVar\ \grt{\grspace}\ \gra{declared-in-this-statement}
		\And
		\grsStatement\ \gra{not-declared-inside-function}
\intertext{%
For \gra{declared-in-this-statement},
a declaration is matched to a reference ($\grC$)
only if no later declaration in this \texttt{var} statement is suitable.
}
	\gra{declared-in-this-statement} &\to
		\grC
		\And
		\gra{ignore-remaining-variables}\ \gra{skip-part-of-this-scope}
		\And \\ &\hspace*{2cm} \And
		\tId\ \tComma\ \gra{not-declared-in-this-statement}
			\\
	\gra{declared-in-this-statement} &\to
		\grC
		\And
		\tId\ \tSemicolon\ \gra{skip-part-of-this-scope}
\end{align*}

It remains to define the rules
for the negative versions
of the three nonterminals in question.
The rules for \gra{this-variable-not-declared}
ensure that the identifier being checked
is different from each of the function's arguments.
\begin{align*}
	\gra{this-variable-not-declared} &\to
		\grCneg
		\And
		\tId\ \tComma\ \gra{this-variable-not-declared}
\intertext{%
Then the search proceeds into the body of the function.
}
	\gra{this-variable-not-declared} &\to
		\grCneg
		\And
		\tId\ \tRightPar\ \tLeftBrace\ \gra{not-declared-inside-function}
			\\
	\gra{this-variable-not-declared} &\to
		\tRightPar\ \tLeftBrace\ \gra{not-declared-inside-function}
\end{align*}

Turning to \gra{not-declared-inside-function},
let $\grsStatementNotVar$ be a new nonterminal
that denotes all well-formed statements
except \texttt{var} statements.
These are the statements
that \gra{not-declared-inside-function}
is allowed to skip without consideration:
whatever declarations are made inside such a statement,
the reference being checked is not in their scope.
\begin{align*}
	\gra{not-declared-inside-function} &\to
		\grsStatementNotVar\ \gra{not-declared-inside-function}
\intertext{%
The rules for $\grsStatementNotVar$
are the same as the rules for $\grsStatement$,
given in Section~\ref{section_the_grammar__statements},
except for the rule for a \texttt{var} statement,
which is not included.
Next, \gra{not-declared-inside-function}
navigates through the nested structure of statements
in the same way as \gra{declared-inside-function}.}
	\gra{not-declared-inside-function} &\to
		\tLeftBrace\ \gra{not-declared-inside-function}
			\\
	\gra{not-declared-inside-function} &\to
		\tIf\ \tLeftPar\ \grsExpr\ \tRightPar\
			\gra{not-declared-inside-function-nested}
			\\
	\gra{not-declared-inside-function} &\to
		\tIf\ \tLeftPar\ \grsExpr\ \tRightPar\ \grsStatement\
			\tElse\ \gra{not-declared-inside-function-nested}
			\\
	\gra{not-declared-inside-function} &\to
		\tWhile\ \tLeftPar\ \grsExpr\ \tRightPar\ \gra{not-declared-inside-function-nested}
			\\
	\gra{not-declared-inside-function-nested} &\to
		\tLeftBrace\ \gra{not-declared-inside-function}
			\\
	\gra{not-declared-inside-function-nested} &\to
		\tIf\ \tLeftPar\ \grsExpr\ \tRightPar\
			\gra{not-declared-inside-function-nested}
			\\
	\gra{not-declared-inside-function-nested} &\to
		\tIf\ \tLeftPar\ \grsExpr\ \tRightPar\ \grsStatement\
			\tElse\ \gra{not-declared-inside-function-nested}
			\\
	\gra{not-declared-inside-function-nested} &\to
		\tWhile\ \tLeftPar\ \grsExpr\ \tRightPar\
			\gra{not-declared-inside-function-nested}
\intertext{%
When a \texttt{var} statement is encountered,
\gra{not-declared-in-this-statement} is invoked to verify
that this statement does not declare the variable under consideration.
At the same time, \gra{not-declared-inside-function}
recursively refers to itself
to make sure that this variable is also not declared
in any subsequent statements.
}
	\gra{not-declared-inside-function} &\to
		\tVar\ \grt{\grspace}\ \gra{not-declared-in-this-statement}
		\And \\ &\hspace*{3cm} \And
		\grsStatementVar \gra{not-declared-inside-function}
\intertext{%
Unlike the nonterminal \gra{declared-inside-function},
which ends the iteration by finding a suitable declaration,
here the iteration ends in a short substring
without variable declarations.
}
	\gra{not-declared-inside-function} &\to
		\anystringwithoutbracesandsemicolons
			\\
	\gra{not-declared-inside-function-nested} &\to
		\anystringwithoutbracesandsemicolons
\end{align*}

Finally, \gra{not-declared-in-this-statement}
applies $\grCneg$ to each identifier
in the current \texttt{var} statement,
in order to ensure that all of them are different
from the identifier in the end of the substring.
\begin{align*}
	\gra{not-declared-in-this-statement} &\to
		\grCneg
		\And
		\tId\ \tComma\ \gra{not-declared-in-this-statement}
			\\
	\gra{not-declared-in-this-statement} &\to
		\grCneg
		\And
		\tId\ \tSemicolon\ \anystringwithoutbraces\ \gra{skip-part-of-this-scope}
\end{align*}

\subsection{Function declarations}\label{section_unambconj__function_declaration}

The rules describing declaration of functions,
given in Section~\ref{section_the_grammar__function_declaration},
suffer from the same kind of problems
as the rules for variable declarations.
This part of the grammar
shall be reconstructed
similarly to what was done
in the above Section~\ref{section_unambconj__variable_declaration}.

First, consider the conjunct
$\grsFunctions\ \gra{this-function-declared-here}$
in one of the rules for \gra{function-declarations},
which concatenates
a prefix with zero or more irrelevant function declarations
to a substring that begins with a declaration of the desired function,
and ends with a call to that function.
This concatenation is ambiguous,
because the function being called
may have multiple declarations,
as demonstrated in the following example.
\begin{equation*}
	\underbrace{\texttt{function() \{ return 0; \} function() \{ return 1; \} main(arg) \{ return function}}_{\gra{function-declarations}}
	\texttt{(); \}}
\end{equation*}
Here the underlined substring 
has two partitions as $\grsFunctions\ \gra{this-function-declared-here}$,
corresponding to the first and the second declaration of \texttt{function()}.

In order to look up functions unambiguously,
instead of using concatenation to get to the desired declaration at once,
one should process all declarations iteratively, one by one,
in the same way as for variable declarations.
This shall be done in a new nonterminal \gra{this-function-declared},
which reimplements the concatenation $\grsFunctions\ \gra{this-function-declared-here}$.
The rules for \gra{this-function-declared}
shall iteratively consider all substrings
that begin with various function declarations
and end with the reference being checked,
and apply \gra{this-function-declared-here} to every such substring.
Furthermore, doing this unambiguously
by the same method as in Section~\ref{section_unambconj__variable_declaration}
requires negative counterparts
of these two nonterminals,
called \gra{this-function-not-declared}
and \gra{this-function-not-declared-here}.

According to this plan,
the three rules for \gra{function-declarations}
dealing with the right parenthesis
are rewritten as follows.
First, if this is a function call,
then the new nonterminal \gra{this-function-declared}
verifies that there is a declaration
of the function being called.
In order to make sure that the rule
indeed deals with a function call
rather than with a declaration,
an extra conjunct states that this prefix of the program
is a sequence of function declarations followed by a header
and an incomplete compound statement,
using the nonterminal $\gra{skip-part-of-this-scope}$
defined in Section~\ref{section_the_grammar__variable_declaration}.
\begin{align*}
	\gra{function-declarations} &\to
		\gra{function-declarations}\ \tRightPar
		\And
		\safeendingstring\, \grsExprCall
		\And \\ &\hspace*{5mm} \And
		\grsFunctions\, \grsFunctionHeader\ \tLeftBrace\ \gra{skip-part-of-this-scope}
		\And
		\gra{this-function-declared}
\intertext{%
Second, if this is a function header,
then the negative version of the new nonterminal (\gra{this-function-not-declared})
shall ensure that there are no earlier declarations
of any functions with the same name
and the same number of arguments.
}
	\gra{function-declarations} &\to
		\gra{function-declarations}\ \tRightPar\
		\And
		\grsFunctions\, \grsFunctionHeader
		\And
		\gra{this-function-not-declared}
\intertext{%
The third case is when the string ending with a right parenthesis
is not of the form ``$\safeendingstring\ \grsExprCall$''.
Since negation is no longer allowed,
one has to list all possibilities
of how a prefix of a well-formed program
could be of such a form;
the following list is inferred from the syntax of this model programming language.
First of all, this prefix may end with an expression
enclosed in brackets;
in this case, the expression
must be preceded by some punctuator character
(which is either an operator or a bracket within a larger expression,
or the last character of some syntactical unit other than an expression)
or by a \texttt{return} keyword.
}
	\gra{function-declarations} &\to
		\gra{function-declarations}\ \tRightPar
		\And
		\anystring\ \anypunctuator\ \WS\ \tLeftPar\ E\ \tRightPar
			\\
	\gra{function-declarations} &\to
		\gra{function-declarations}\ \tRightPar
		\And
		\safeendingstring\ \tReturn\ \tLeftPar\ E\ \tRightPar
\intertext{%
The remaining possibility
is that the right parenthesis under consideration
is a part of an \texttt{if} or a \texttt{while} statement.
}
	\gra{function-declarations} &\to
		\gra{function-declarations}\ \tRightPar
		\And
		\safeendingstring\ \tIf\ \tLeftPar\ E\ \tRightPar
			\\
	\gra{function-declarations} &\to
		\gra{function-declarations}\ \tRightPar
		\And
		\safeendingstring\ \tWhile\ \tLeftPar\ E\ \tRightPar
\end{align*}

The next goal is to define the rules for the new nonterminals
\gra{this-function-declared} and \gra{this-function-not-declared}.
The first rule for \gra{this-function-declared}
skips any function declarations,
as long as there is still a declaration of this function later on.
\begin{align*}
	\gra{this-function-declared} &\to
		\grsFunction\ \gra{this-function-declared}
\intertext{%
The second rule describes a substring
that begins with a desired function declaration
and ends with a reference to the same function:
this condition is checked by the nonterminal \gra{this-function-declared-here},
using the rules defined in Section~\ref{section_the_grammar__function_declaration}.
}
	\gra{this-function-declared} &\to
		\gra{this-function-declared-here}
		\And
		\grsFunction\ \gra{this-function-not-declared}
\intertext{%
In order to keep the grammar unambiguous,
the second conjunct of this rule
ensures that no later function declaration
matches this reference.
The last rule handles a special case,
where the substring contains a function header
and a part of the same function's body
ending with its recursive call to itself.
}
	\gra{this-function-declared} &\to
		\gra{this-function-declared-here}
		\And
		\grsFunctionHeader\ \tLeftBrace\ \gra{skip-part-of-this-scope}
\end{align*}
Here the nonterminal \gra{skip-part-of-this-scope},
reused from Section~\ref{section_the_grammar__variable_declaration},
ensures that the substring contains no other function declarations.

Now consider checking a function
for having no declaration
(\gra{this-function-not-declared}).
This is done for strings of the same form
as for \gra{this-function-not-declared}:
that is, for a substring
beginning with zero or more function declarations,
and continued either with a function header
or with an incomplete function declaration ending with a function call.
One has to verify that the function in the end of the substring
has no earlier declarations.
For substrings that begin with a function declaration,
the first rule states
that the function is declared neither here, nor later.
\begin{align*}
	\gra{this-function-not-declared} &\to
 		\gra{this-function-not-declared-here}
		\And
		\grsFunction\ \gra{this-function-not-declared}
\intertext{%
Once all earlier function declarations
are processed in this way,
eventually there is one of two base cases to handle.
First, \gra{this-function-not-declared}
may have to deal with a substring
that consists of just one function header.
It has already been checked that this function
has no earlier declarations,
and the iteration ends here.
}
	\gra{this-function-not-declared} &\to
		\grsFunctionHeader
\intertext{%
The second case is when a substring
begins with a function header
and continues with an incomplete function body
ending with a function call expression.
Then, the following rule ensures
that this is \emph{not} a valid recursive call to itself.
}
	\gra{this-function-not-declared} &\to
		\gra{this-function-not-declared-here}
		\And
		\grsFunctionHeader\ \tLeftBrace\ \gra{skip-part-of-this-scope}
\end{align*}

Finally, it remains to write down the rules
for \gra{this-function-not-declared-here}.
Dually to the rule for \gra{this-function-declared-here}
from Section~\ref{section_the_grammar__function_declaration},
here one can say that either the two functions have different names,
or they have the same name but a different number of arguments.
\begin{align*}
	\gra{this-function-not-declared-here} &\to
		\gra{different-function-name}
			\\
	\gra{this-function-not-declared-here} &\to
		\gra{same-function-name}
		\And
		\gra{different-number-of-arguments}
\end{align*}
Name mismatch is tested using $\grCneg$.
\begin{align*}
	\gra{different-function-name} &\to
		\grCneg\ \tLeftPar\ \grsListOfExpr\ \tRightPar
\intertext{%
Mismatch in the number of arguments
is described by the same kind of rules
as for match (\gra{n-of-arg-equal}, \gra{n-of-arg-equal-0}).
}
	\gra{different-number-of-arguments} &\to
		\tId\ \tLeftPar\ \gra{n-of-arg-less}\ \tRightPar
		\ | \
		\tId\ \tLeftPar\ \gra{n-of-arg-greater}\ \tRightPar
			\\
	\gra{different-number-of-arguments} &\to
		\tId\ \tLeftPar\ \gra{n-of-arg-less-0}\ \tRightPar
		\ | \
		\tId\ \tLeftPar\ \gra{n-of-arg-greater-0}\ \tRightPar
			\\
	\gra{n-of-arg-less} &\to
		\gra{n-of-arg-less}\ \tComma\ E
		\ \big| \
		\gra{n-of-arg-equal}\ \tComma\ E
			\\
	\gra{n-of-arg-greater} &\to
		\tId\ \tComma\ \gra{n-of-arg-greater}
		\ \big| \
		\tId\ \tComma\ \gra{n-of-arg-equal}
			\\
	\gra{n-of-arg-less-0} &\to
		\gra{n-of-arg-less-0}\ \tComma\ E
		\ \big| \
		\gra{n-of-arg-equal-0}\ E
			\\
	\gra{n-of-arg-greater-0} &\to
		\tId\ \tComma\ \gra{n-of-arg-greater-0}
		\ \big| \
		\tId\ \gra{n-of-arg-equal-0}
\end{align*}

\subsection{The main function}\label{section_unambconj__main_function}

The rules for the main function
defined in Section~\ref{section_the_grammar__programs}
are ambiguous,
because the program may contain multiple declarations for the main function.
To be precise, the concatenation
$\grsFunctions\ \grsFunctionMain\ \grsFunctions$
in the rule for $\grsProgram$
is ambiguous,
because there are as many partitions
as there are main functions declared in a program
(which is obviously ill-formed).

The rule for $\grsProgram$ is therefore rewritten
by using a new nonterminal symbol $\grsFunctionsWithMain$
that represents any sequence of function declarations
that contains a main function.
\begin{align*}
	\grsProgram &\to
		\WS\ \grsFunctionsWithMain
		\And
		\WS\ \gra{function-declarations}
\intertext{%
The rules for $\grsFunctionsWithMain$
add function declarations, one by one,
as long as they are \emph{not} for a main function.
Once the last declaration of a main function in the program is located,
it does not matter which functions are defined before it.
}
	\grsFunctionsWithMain &\to
		\grsFunctionsWithMain\ \grsFunctionNotMain \ | \
		\grsFunctions\ \grsFunctionMain
			\\
\intertext{%
(note that if the first rule were replaced with
$\grsFunctionsWithMain \to \grsFunctionsWithMain\ \grsFunction$,
then the grammar would become ambiguous again)
Finally, it remains to define
a declaration of a function
that is not a main function,
and to do this without using the negation.
The first possibility is
that the function's name is not \texttt{main}.
Assume that the set of all identifiers other than \texttt{main}
is defined in a new nonterminal symbol $\tIdNotMain$,
with its rules simulating a finite automaton.
}
	\grsFunctionNotMain &\to
		\tIdNotMain\ \WS\ \tLeftPar\ \grsListOfDistinctIds\ \tRightPar\ \grsStatementCompoundRet
		\And
		\tId\ \tLeftPar\ \gra{all-variables-declared}
			\\
	\grsFunctionNotMain &\to
		\tIdNotMain\ \WS\ \tLeftPar\ \tRightPar\ \grsStatementCompoundRet
		\And
		\tId\ \tLeftPar\ \gra{all-variables-declared}
\intertext{%
The other case is when the function is called \texttt{main},
but has none or at least two arguments.
}
	\grsFunctionNotMain &\to
		\grt{m}\ \grt{a}\ \grt{i}\ \grt{n}\ \WS\ \tLeftPar\ \grsListOfDistinctIdsTwoPlus\ \tRightPar\ \grsStatementCompoundRet
		\And
		\tId\ \tLeftPar\ \gra{all-variables-declared}
			\\
	\grsFunctionNotMain &\to
		\grt{m}\ \grt{a}\ \grt{i}\ \grt{n}\ \WS\ \tLeftPar\ \tRightPar\ \grsStatementCompoundRet
		\And
		\tId\ \tLeftPar\ \gra{all-variables-declared}
\end{align*}
The rules defining $\grsListOfDistinctIdsTwoPlus$
are a variant of those for $\grsListOfDistinctIds$.

This completes the last correction to the grammar.

\begin{proposition}
The set of well-formed programs
in the model programming language
is described by an unambiguous conjunctive grammar
with 187 nonterminal symbols
and 3828 rules.
\end{proposition}

There are so many rules
because of the two finite automaton simulations
for $\tId$ and for $\tIdNotMain$,
and also because of the many rules
of the form $\Citerateneg \to \Cc{$\sigma$} \tau$
needed to compare identifiers for inequality.

Even though the size of the grammar has increased ten-fold
in comparison with the original ambiguous Boolean grammar,
the overall impact on the worst-case parsing complexity is beneficial.
Indeed, for unambiguous conjunctive grammars,
as well as for unambiguous Boolean grammars,
the theoretical upper bound on their parsing complexity
is $O(n^2)$, where $n$ is the length of the program~\cite{BooleanUnambiguous}.
This is a significant improvement over cubic-time
or slightly better performance
of the existing algorithms on highly ambiguous grammars.
In particular, the GLR parsing algorithm for Boolean grammars~\cite{BooleanLR}
works in square time on any unambiguous conjunctive grammar
(even though no formal proof of that fact has ever been given).
On some inputs, such as on the following program,
the GLR parser for the grammar from Section~\ref{section_the_grammar}
is forced into cubic-time behaviour,
whereas the unambiguous GLR parser works significantly faster
due to its guaranteed square-time performance.
\begin{equation*}
	\texttt{main(x) \{ return}\
	\underbrace{\texttt{x+x+ \ldots +x}}_n
	\texttt{; \}}
\end{equation*}
The grammar-dependent constant factor in the parsing complexity
is small in comparison,
as it is only linear in the size of a GLR parser's table.

No proof that the given grammar is unambiguous is given in this paper,
just like there is no proof that it describes
exactly the desired programming language.
Although it is possible to prove this kind of results by hand,
for such a large grammar,
that would hardly be practical.
For ordinary grammars (``context-free'' according to Chomsky),
there exists several automated methods for ambiguity detection~\cite{Basten,Schmitz}.
Investigating any methods of this kind
for conjunctive and Boolean grammars
is one of the many open problems
suggested by this work.

\subsection{A possible linear conjunctive grammar}

It is possible that the grammar presented in this section
could be further reconstructed
into a linear conjunctive grammar.
This is a simpler model
notable for its equivalence
to a family of cellular automata~\cite{LinearAutomata}.
The best known complexity upper bound
for linear conjunctive grammars
is still $O(n^2)$,
so this transformation
is of a purely theoretical interest.

Some parts of the grammar,
such as identifier comparison in Section~\ref{section_the_grammar__identifier},
are already almost linear conjunctive,
and require only very obvious transformations.
Another kind of inessential non-linearity
is caused by the nonterminal symbols representing tokens,
which are being concatenated throughout the grammar.
Since each of them defines a regular language,
those languages can be directly expressed within linear rules.

The rules defining the nested structure
of expressions, statements and functions,
given in Sections~\ref{section_the_grammar__expressions}--\ref{section_the_grammar__functions},
are essentially non-linear.
However, that structure is simple enough to be recognized
by \emph{input-driven pushdown automata}~\cite{Mehlhorn},
also known under the name of \emph{visibly pushdown automata}~\cite{AlurMadhusudan}---%
and those automata can in turn be simulated
by linear conjunctive grammars~\cite{linconj_vs_dcfl}.

The rules concerned with declaration before use,
in their final form
given in Sections~\ref{section_unambconj__variable_declaration}--%
\ref{section_unambconj__function_declaration},
have many essentially non-linear concatenations.
As a representative example,
consider the rule
	$\gra{not-declared-inside-function} \to
		\tIf\ \tLeftPar\ \grsExpr\ \tRightPar\
		\tLeftBrace\ \gra{not-declared-inside-function-nested}$.
Strictly speaking, it concatenates 6 nonterminal symbols.
However, four of them represent tokens
and accordingly define only regular sets,
whereas $\grsExpr$ is a bracketed construction
recognized by an input-driven pushdown automaton.
It is conjectured that this and all other such concatenations in the grammar
can be simulated by linear conjunctive rules
by further elaborating the known simulation
of input-driven pushdown automata~\cite{linconj_vs_dcfl}.

If the suggested transformation works out,
the resulting linear conjunctive grammar
will be large and incomprehensible.
It should then be regarded as a kind of ``machine code''
that implements the grammar
given in Sections~\ref{section_the_grammar__alphabet}--\ref{section_unambconj__main_function}.

\section{Afterthoughts}\label{section_afterthoughts}

The main purpose of this paper
was to demonstrate that the expressive power
of conjunctive and Boolean grammars
is sufficient to describe
some of the harder syntactic elements
of programming languages.
This becomes the first successful experience
of constructing a complete formal grammar
from a practically parsable class
for a programming language.

As a first experience, it has many shortcomings.
The model programming language is quite restricted,
and enriching its syntax even a little bit,
while staying within Boolean grammars,
would be challenging, if at all possible:
for instance, introducing type checking would be problematic,
and it is quite possible that doing this
would already be beyond the power of Boolean grammars.
Some parts of the resulting description are quite natural,
some are admittedly awkward.
Grammar maintenance would be difficult,
as a small change in the syntax of the language---%
such as, for instance, changing the punctuators used in a single language structure---%
may require rewriting other parts of the grammar in a non-trivial way.
Square-time parsing may be fast enough for some applications,
but it is still too slow
in comparison with standard parsers for programming languages.
These observations
suggest the following research directions.

First,
could there exist \textbf{a substantially faster parsing algorithm}
that would still be applicable
to (some variant of) the grammar given in this paper?
The existing linear-time algorithms
for subclasses of conjunctive grammars~\cite{AizikowitzKaminski_LR0,BooleanLL}
are too restrictive.
If the model programming language defined here
were parsed by a standard human-written program,
that parser would maintain a symbol table of some sort,
filling in new entries upon reading declarations,
and looking it up for every reference.
This suggests that if a prospective parsing algorithm
is to parse this language
substantially faster than in square time,
then it will most likely need some advanced data structures.
Would it be possible to adapt the GLR algorithm
to index its graph-structured stack
using a symbol table?
The grammar in Section~\ref{section_unambconj}
demonstrates the kind of rules
such an algorithm is expected to handle.

The second research direction
is motivated 
by the awkward parts of the grammar
and by the limitations of the model programming language.
These can be regarded
as signs of imperfection of conjunctive and Boolean grammars,
and from this perspective,
the goal is to find \textbf{a new grammar formalism},
in which all that could be done better.
Any such formalisms must maintain efficient parsing algorithms,
and would likely be found
by further extending conjunctive or Boolean grammars.
Over the years this paper has been under preparation,
Barash~\cite{BarashProgramming} found out
that the grammar in this paper can be improved
and the model language extended,
if the grammar model is augmented
with \emph{operators for referring to the contexts} of a substring.
The resulting new grammar model
has a purely logical definition
and still has a cubic-time parsing algorithm~\cite{GrammarsWithContexts}.
It would be interesting to see any other new attempts
to define a suitable grammar model.

The last suggested topic concerns
conjunctive and Boolean grammars as they are,
and their \textbf{applications}.
The grammar constructed in this paper
is an extreme case of using these models,
meant to demonstrate the possibility
of describing a number of constructs.
It remains to find more practical ways of using
these grammar construction methods.

For instance, it could be possible
to define a slightly higher-level notation for syntax
based on conjunctive and Boolean grammars.
In the grammar presented in this paper,
handling some simple issues of lexical analysis
turned out to be quite inconvenient,
because every rule of the grammar
had to define the splitting of a string into tokens
and implement the \emph{longest match} principle.
If a traditional lexical analyzer could prepare a program for parsing---%
for example, by ensuring that there is exactly one space character between every two tokens---%
then a Boolean grammar may describe only the well-tokenized inputs,
and this would simplify the definition.
Since the family of formal languages defined by Boolean grammars
is known to be closed under inverse finite transductions~\cite{BooleanInvGSM},
there are no theoretical obstacles
to combining a finite transducer and a Boolean grammar in this way.
The question is to design a convenient syntactic formalism along these lines.

In general, conjunctive and Boolean grammars
should not be difficult to implement in existing projects,
since the widely used GLR parsing algorithm
can handle conjunctive and Boolean rules~\cite{Megacz,BooleanLR}.
Besides parsing algorithms,
no other essential grammar technologies and tools
have yet been developed for Boolean grammars.
Grammar maintenance, amgibuity detection, attribute evaluation, etc.---%
none of these subjects have yet been addressed.
Investigating these problems,
perhaps finding out some ways
to extend the ideas from ordinary grammars
to grammars with Boolean operations,
is another research direction.

\end{document}